\documentstyle[aps,prb,epsfig,twocolumn,floats,psfig]{revtex}
\advance\textheight by 0.14in
\advance\topmargin by -0.1in
\begin{document}
\draft
\twocolumn[\hsize\textwidth\columnwidth\hsize\csname @twocolumnfalse\endcsname
\widetext
\title{Vortex Deconfinement in the $XY$ Model with a Magnetic Field}
\author{H.~A. Fertig$^{(1)\!}$ and Kingshuk Majumdar$^{(2)\!}$}
\address{$^{(1)\!}$ Department of Physics and Astronomy, University of Kentucky, Lexington, KY
40506-0055;\\ 
$^{(2)\!}$ Department of Physics, Berea College, Berea, KY 40404}
\date{\today}
\maketitle

\begin{abstract}
{We study vortex unbinding for the classical two-dimensional $XY$ model in a
magnetic field on square and triangular lattices. A renormalization group analysis
combined with duality in the model shows that at high temperature and high field,
the vortices unbind as the magnetic field is lowered in a two-step process:  strings
of overturned spins first proliferate and then vortices unbind. The transitions are highly continuous
but are not of the Kosterlitz-Thouless type. The unbound vortex fixed point is shown to inherit
properties of the underlying lattice, in particular containing a set of nodal lines that reflect the
lattice symmetry.}
\end{abstract}
\bigskip
]
\narrowtext

\section{Introduction}
\label{Intro}
Topological defects play a crucial role in a host of phenomena in condensed
matter systems~\cite{nelson_book}. Among these are Kosterlitz-Thouless (KT)
transitions~\cite{thouless}, in which vortex-like defects in a two dimensional
system unbind above some
critical temperature. Kosterlitz-Thouless transitions are important for
understanding disordering in planar ($XY$) magnets, destruction of
superfluidity and superconductivity in thin films, melting of two dimensional crystals,
and transitions among different liquid crystal states~\cite{nelson_book}. 
Vortex unbinding is related to the roughening transition, in which the
interface between two materials fluctuates without bound above a
critical temperature~\cite{chaikin}. 
They are 
important for understanding ground state properties of 1+1 dimensional
quantum systems (Luttinger liquids)~\cite{gogolin}.  They have also
recently found relevance in understanding the states of stripe systems
in high temperature superconductors and quantum Hall 
systems~\cite{kivandfrad,fertig_str}.  The
$XY$ ferromagnet is often considered the paradigm of all these systems,
largely because of the (somewhat misleading) simplicity of its Hamiltonian.

Much less attention has been paid to what becomes of the vortex unbinding
transition in the presence of a symmetry-breaking field, such as a magnetic
field tending to align the $XY$ spins along some particular direction,
in spite of the fact that most of the systems discussed above have realizations
in which such symmetry-breaking fields are present.  One major reason
for this is that the standard KT analysis breaks down under the most 
interesting circumstances when a symmetry breaking field is present.
Usually, one assesses whether vortices may be incorporated into a 
simple theoretical description of a system by renormalization of
the parameters of the Hamiltonian.  This is possible when the vortices
are bound tightly into pairs of vanishing net vorticity.  In the 
renormalization group (RG) approach to the KT transition, the paired
state is signaled by a flow in which the vortex fugacity, $e^{-E_c/k_BT}
\equiv y/2$,
with $E_c$ the core energy of a vortex, vanishes.  The interpretation
of this is that an effective long-wavelength theory of the system may
be constructed with vanishing vortex density ($y=0$); the vortex
degrees of freedom are {\it irrelevant} in the renormalization group sense~\cite{cardy}.

For small values of $y$ and above the Kosterlitz-Thouless transition temperature
$T_{KT}$, one finds that $y$ grows rather than shrinks, signaling the
relevance of vortices in the system, so that a theory without
vortices cannot be qualitatively correct.  This sharp change
in behavior is accompanied by an essential singularity in the free-energy,
so that vortex unbinding constitutes a continuous phase transition.
Because the RG flows contain a set of simple attractive fixed points
 (the Hamiltonians with $y=0$ to which
the low temperature microscopic Hamiltonians flow under the RG),
thermodynamic properties of the transition may be 
computed~\cite{nelson_book,thouless}.

For many systems with symmetry-breaking fields, the existence of this set
of stable fixed points is lost~\cite{jose}.  A prototype of this -- and the
system we focus on in this paper -- is the $XY$ system with a magnetic
field. It is defined by the Hamiltonian
\begin{equation}
{\cal H}_{XY}/k_B T= -K \sum_{<{\bf r},{\bf r^{\prime}}>} \cos(\theta_{\bf r}-\theta_{\bf r^{\prime}}) - 
h\sum_{\bf r} \cos{\theta_{\bf r}}.
\label{XYmodel}
\end{equation}
In Eq.~\ref{XYmodel}, $\theta_{\bf r}$ represents the angle of a
planar spin at location ${\bf r}$, $K$ is an exchange coupling between
nearest neighbor spins, and $h$ is the magnitude of an effective
magnetic field tending to align the spins along the ${\bf \hat{x}}$ axis.
For convenience from now on we will choose our units of energy so that
$k_BT=1$. 
A perturbative RG analysis\cite{jose} when applied to this
system shows that either the symmetry-breaking term in the Hamiltonian
or the vortex fugacity, or both, are relevant even when they
are initially very small.  The absence of a stable fixed point means this
perturbative analysis cannot provide us with a simple Hamiltonian to
describe the state of the system.  This has been interpreted to mean
that there is no phase transition, and presumably no possibility of
vortex unbinding, when a symmetry-breaking field is present (see,
for example, Chap. 2 of Ref.~\onlinecite{nelson_book}). In this paper,
we will describe a new analysis that suggests that while the first part
of this interpretation (no phase transition) is at some level true, the
second part (no vortex unbinding) is not: we will develop a RG
description that allows for vortex unbinding, but with a fixed point
structure that remarkably avoids the free energy singularity usually
associated with a phase transition.  Some of the results discussed here
were described elsewhere\cite{herb1}.  In this paper we provide details
of those calculations, and extend them to show how the underlying lattice symmetry
may be incorporated into the model.  

An important perspective on why unbinding in the 
$XY$ ferromagnet is different in the presence of a magnetic
field than without it comes from examining the interaction of a single vortex-antivortex
pair.  Without symmetry breaking, this is
well known to be a slow, logarithmic potential with respect to
pair separation.  This arises because the lowest energy spin
configuration, subject to the constraint that the system contains
a vortex and an antivortex, approaches the ferromagnetic ground state
slowly as one moves away from the pair.  Such a configuration
is energetically expensive for $h>0$  since
it involves many spins tilted away from the direction specified by the
field; the system instead forms a {\it string} of overturned spins
connecting the vortex-antivortex pair\cite{girv}.  The rotation of the spins
through $2\pi$ as one moves through the string is essentially 
identical to a kink soliton of the sine-Gordon model\cite{raja},
carrying a characteristic energy per unit length proportional to 
$\sqrt{Kh}$.  Thus, the potential energy for 
a pair of vortices increases {\it linearly} with separation.
Such linear potentials arise in the context of strong interactions,
and leads to confinement\cite{quark}, a situation in which individual
particles (quarks for strong interactions, vortices in the $XY$ model)
do not appear in isolation.  From this analogy one might conclude there
is no unbound vortex state of this system.

The difficulty with this argument is that it ignores the statistical mechanics
of the string connecting the vortices.  At high temperatures, one must
include entropy in the free energy of the string connecting a single vortex-antivortex 
pair.  The number of configurations for a string grows 
exponentially\cite{vanderzande} with its length $L$, so that at high
temperature a string cannot bind a single vortex-antivortex pair.
Moreover, the string tension -- its energy per unit length -- is
renormalized downward by small breaks  (vortex-antivortex pairs),
which may occur along its length.  Finally, at high temperature
the $XY$ system will be flooded with vortices and antivortices,
so that the distance between neighboring pairs will be of the
same order as the pair sizes themselves.  This situation is highly
reminiscent of what is expected to occur in a quark-gluon plasma,
in which quarks are deconfined\cite{quark}. 
Deconfinement at high temperature is clearly also a possibility for vortices
in the $XY$ system.   To decide if and when this occurs, we will develop
a model that deals {\it directly} with the strings, and contains a parameter
which we will identify with the string tension.  This  string tension
can be driven to zero when renormalized by fluctuations -- signaling
an unbinding transition for the vortices.

Before beginning our technical discussion, we summarize our results.
We develop an effective model for this system, focusing on a
representation in which the string degrees of freedom are explicit.
The model has a dual form, 
which may be interpreted as a solid-on-solid model
with screw dislocations.  The resulting low energy excitations 
contain both open and closed domain walls.  The endpoints of the
domain walls -- the screw dislocations -- are degrees of freedom
dual to the vortices.  We initially focus on the states of these
dislocations, and then  use the duality to draw conclusions
about the allowed states of the vortices.
The perturbative RG analysis we follow is valid in
the large $h$ and small $y$ limit.
It generates a
parameter $\rho$, which represents the energy per unit distance
for separating two endpoints of a single open domain wall, keeping
the length of the domain wall fixed.  The RG equations indicate that
$\rho$ may flow to zero, or may flow to a finite value; we interpret
the latter as a bound dislocation state.  
The $\rho=0$ fixed point is accessible when the domain walls
are rough, i.e., when they are unbounded in size and
percolate through the system.   Since arbitrarily large domain
walls may be cut open with endpoints as far apart as we 
like without extra energy cost at the $\rho=0$ fixed point,
we identify this as an unbound dislocation phase.

In the RG flows,
the $\rho=0$ fixed point appears at the end of
a line of fixed points with $\rho \ge 0$, so that the flows 
accumulate at the $\rho=0$ for the unbound phase.  The string tension
$\rho$ grows continuously from zero when the microscopic
parameters of the Hamiltonian cross into the values corresponding
to the bound dislocation phase. For fixed $K$ and $E_c$,
bound dislocations occur for $h$ smaller than some critical
value.  

An important observation is that {\it there is no relevant
direction in the parameter space of the Hamiltonian leading away from the 
unbound dislocation fixed point.}  This is a remarkable result: in
the absence of a relevant direction, there is no mechanism by which
the effective free energy can accumulate a singularity in the RG as
one integrates out short distance scales.  With no free energy singularity,
one does not expect to find singularities in any thermodynamic
quantities for the system.  Nevertheless, we can sharply distinguish
the bound and unbound vortex phases, for example by measuring
the diffusion constant for vortices, or (equivalently) measuring fluctuations
in the vortex dipole moment of the system.  Recent Langevin dynamics
simulations\cite{herb2} focusing on the latter quantity have confirmed
the basic results of the RG studies presented here; we will discuss these
in more detail below.   The important lesson at this point is that
vortex deconfinement in this system is not a phase transition in
the usual sense.  One can distinguish the bound and unbound vortex
phases through transport properties, correlation functions, or statistics
of specific fluctuations in the system.  But they cannot be distinguished by
qualitative differences in quantities that may be expressed as derivatives
of the free energy; i.e., thermodynamic quantities.  The possibility of 
such unusual transitions has been noted in rigorous studies of phase
transitions\cite{ruelle}, but to our knowledge this is the first concrete
example of such behavior.

The unbound dislocation fixed point, we shall see, has a unique signature:
it contains one or more lines of zero energy modes in the Brillouin zone.
Our earlier analysis\cite{herb1} focused on the case of a single such nodal
line. This breaks the discrete rotational symmetry of the
lattice, and we will see it arises from using a particular choice of gauge
which leads to approximations breaking the symmetry.  We will show that
the underlying symmetry of the lattice may be retained without changing
any of the qualitative results.  In the case of a square lattice, the fixed point
contains two nodal lines, and for a triangular lattice it contains three.
The number of nodal lines is determined by the number of independent
directions a domain wall may exit from a dislocation;
it is essentially the number of distinguishably different types of dislocations
the system supports.  We note that these fixed points
might be regarded as a classical analog of similar Hamiltonians arising in
the context of quantum spin models with ring exchange~\cite{fisher}.

To understand what all this implies about vortex unbinding, we must
re-express the partition function in its dual form.  The dual Hamiltonian
we will see
is identical to the dislocation Hamiltonian, provided we exchange 
$E_c \leftrightarrow 1/2h$ and substitute $K \rightarrow 1/4\pi^2K$.
Since the unbound dislocation phase occurs for large $E_c$ and small
$1/2h$,  we expect unbound vortices can be found at large $1/2h$ and
small $E_c$.  That an unbound vortex phase does not occur for large
values of $E_c$ shows how different the transition is from the KT
phenomenology, and also is consistent with the failure of perturbative
RG calculations\cite{jose} with small 
$y=e^{-E_c}$ (i.e., {\it large} $E_c$) to capture the transition.
Since dislocation unbinding occurs at a finite value of $1/2h$,  dislocation
unbinding and vortex unbinding {\it cannot} be the same transition.
This tells us there must be an intermediate phase in which both 
dislocations and vortices are bound.  The nature of this phase
follows from the observation that domain walls are rough even in
the bound dislocation phase.  At long wavelengths, we can 
ignore the dislocations, so one is effectively in the rough (high temperature)
phase of a solid-on-solid model.  This is dual to a bound (low temperature)
phase of logarithmically interacting vortices~\cite{chaikin}, so we identify
the large $E_c$, $1/2h$ limit of the parameters as a logarithmically bound
dislocation phase.  From the duality, we expect this implies the vortices will
also be logarithmically bound in this region of parameters.  Physically,
we can understand this by recognizing the domain walls in the dislocation
representation are dual to the strings in the vortex representation, and both
are proliferated for these parameters.  Proliferated strings do not linearly
confine the vortices, but they can (and do) upwardly renormalize the logarithmic
interactions between them, enough so that the vortices may be bound together
even if $T>T_{KT}$.

Finally, one may ask what the unbound dislocation phase looks like when
expressed in terms of the vortex degrees of freedom.  Since this occurs
for very large values of $h$ and $E_c$, the bare string tension between vortices
is very large and the vortices are dilute.  In this case it is clear that the vortices
are linearly confined.  Thus we identify the highest temperature phase of
the dislocations -- a deconfined phase -- with the lowest temperature phase of
the vortices -- a linearly confined phase.  

\begin{figure}[tb]
\begin{center}
 \vbox to 5.0cm {\vss\hbox to 10cm
 {\hss\
   {\includegraphics{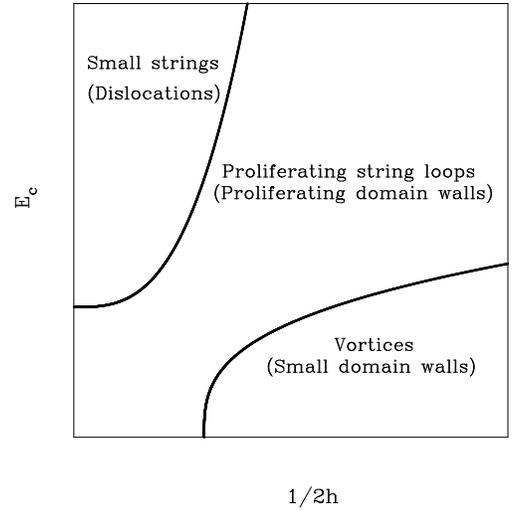}
   }
  \hss}
 }
 \end{center}
\vspace{1.8cm}
\protect \caption{Schematic phase diagram for $XY$ model in a magnetic field
for $K \approx 1/2\pi$.  Upper left corner
represents linearly confined phase (unbound screw dislocations in dual representation).  
Middle phase contains proliferated
loops in both descriptions and vortices (dislocations) are logarithmically bound. 
Lower right corner contains unbound vortices 
(linearly confined dislocations).}
\protect \label{pdsimple}
\end{figure}

All these considerations suggest that for large $E_c$ and/or small $h$, there
are three phases for the vortices: a linearly confined phase, a logarithmically
confined phase, and a deconfined phase.  This is illustrated in Fig.~\ref{pdsimple},
which is the simplest phase diagram one may draw consistent with
the perturbative RG analysis. 

We are led to a picture in
which vortices deconfine in a two-step process: as fluctuations increase,
the strings connecting them proliferate, but vortex-antivortex pairs remain bound
by a residual logarithmic attraction; at still higher levels of fluctuations
the closed string loops may break open, and the vortices deconfine.
We note that simulations strongly support the existence of two different
bound phases, as well as the unbound vortex phase\cite{herb2}.

This article is organized as follows.  We begin  Section~\ref{Model} by expressing the 
system as a Villain model, and then  formulate the model
as a continuum theory.  In Section~\ref{RGanalysis} we develop the RG calculations and show how the
unusual fixed point structure emerges.  Section~\ref{fullsymmetry} discusses more
careful formulations that respect the lattice symmetry for square and
triangular lattices.  Section~\ref{discussion} contains discussion of the numerical 
integration of the RG equations, and we conclude with a summary in Section~\ref{summary}.
Two Appendices provide further details of the calculations.

\section{Derivation and Interpretation of the Model}
\label{Model}
\subsection{Representation in Terms of Domain Walls and Screw Dislocations}
In this section, we present the derivation of the model we will be considering.
Many of the steps we take may be found in Ref.~\onlinecite{jose}, and indeed
our final result is equivalent to theirs, although the representation on which
we finally perform our RG analysis is considerably different, leading to
different results.  In any case, for completeness we present the derivation
in its entirety.

The partition function for the $XY$ model may be expressed as 
${\cal Z}_{XY}=\int {\cal D}\theta e^{-{\cal H}_{XY}}$.   Because of the
cosines appearing in the exponent, it is very difficult to make progress
working directly with this Hamiltonian.  Since we are interested in
phases and long wavelength properties of the system,
we can replace the model with any other that
contains the same symmetries and retain the correct physics.
A particularly useful model to adopt is the Villain
model~\cite{villain}, in which one makes the replacement
\begin{equation}
e^{C\cos \theta} \rightarrow \sum_{m=-\infty}^{\infty}e^{-C(\theta - 2\pi m)^2/2}
\label{poisson}
\end{equation} 
whenever a cosine appears in the exponent.  $m$ above is an integer
variable, and the important observation is that the weight as a function of
$\theta$ has period $2\pi$ on both sides of Eq.~\ref{poisson}.
The replacement of a cosine in the exponent with a quadratic form
allows us to make some progress.
Applying
this to ${\cal Z}_{XY}$, we are led to the replacement 
\begin{eqnarray}
{\cal Z}_{XY} \rightarrow {\cal Z}_{VM}\:\:\:\:\:\:\:\:\:\:\:\:\:&& \cr
=\sum_{m_{{\bf r},{\bf r^{\prime}}}}
\sum_{n_{\bf r}} \int {\cal D}\theta 
\exp &\Biggl[& -{K \over 2}\sum_{\langle {\bf r},{\bf r'} \rangle}
(\theta_{\bf r} - \theta_{\bf r'}-2\pi m_{{\bf r},{\bf r'}})^2 \cr
&-&{h \over 2} (\theta_{\bf r} - 2\pi n_{\bf r})^2 \Biggr].
\label{trans1}
\end{eqnarray}
The functional integral over the $\theta$ field may be performed with
the help of the Poisson resummation formula.  This states that for a function
$g(m)$ summed over an integer field $m$,
$$
\sum_{m=-\infty}^{\infty} g(m) =\sum_ {n=-\infty}^{\infty}
\int_{-\infty}^{\infty} d\phi \;g(\phi) e^{-2\pi i n\phi}.
$$ 
Applying this to both integer fields in Eq.~\ref{trans1} and shifting variables, we
arrive at the expression
\begin{eqnarray}
{\cal Z}_{VM} &=&\sum_{S_{{\bf r}{\bf r^{\prime}}}}
\sum_{T_{\bf r}} \int {\cal D}\theta \int {\cal D} \phi  \int {\cal D} \psi \cr
\times \exp &\Bigl[& -2\pi^2 K \sum_{\langle {\bf r},{\bf r'} \rangle}\phi_{ {\bf r},{\bf r'} }^2
-2\pi^2 h\sum_{\bf r} \psi_{\bf r}^2 \cr
&-&2\pi i \sum_{\langle {\bf r},{\bf r'} \rangle} S_{{\bf r}{\bf r^{\prime}}}\phi_{{\bf r},{\bf r'} }
\Bigr]  \cr
\times \exp &\Bigl[& -2\pi i \sum_{\bf r} T_{\bf r} \psi_{\bf r}
 -i \sum_{\langle {\bf r},{\bf r'} \rangle}(\theta_{\bf r} - \theta_{\bf r'})S_{{\bf r}{\bf r^{\prime}}} \cr
&-&i \sum_{\bf r}\theta_{\bf r} T_{\bf r}
\Bigr],
\label{trans2}
\end{eqnarray}
where $ S_{{\bf r}{\bf r^{\prime}}}$ and $T_{\bf r}$ are the integer
fields arising from the Poisson resummations.
The continuous fields in Eq.~\ref{trans2} may now all be integrated out.
We focus first on the angular variables $\theta$, for which the integration
produces a product of $\delta$-functions.  These may be conveniently written
if we re-express the bond integer field $ S_{{\bf r}{\bf r^{\prime}}}$
as a vector field ${\bf S}({\bf r})$, with components 
$ S_i({\bf r}) = S_{{\bf r},{\bf r}+{\bf \Delta}_i}$,
where ${\bf \Delta}_i$ are vectors denoting nearest
neighbor bonds in the ``positive'' direction  (For example, on a square
lattice one may take ${\bf \Delta}_1=a_0 {\bf \hat{y}}$, ${\bf \Delta}_2=a_0 {\bf \hat{x}}$.).
It is important to note that the dimensionality of  ${\bf S}$ is {\it not}
set by the (two-) dimensionality of the lattice, but rather is given by (half) the
coordination number; for example, in the triangular lattice discussed below,
${\bf S}$ is a three-dimensional vector field.  Whatever its dimensionality,
we can define a divergence,
$$
{\bf \nabla} \cdot {\bf S}({\bf r}) \equiv \sum_i \bigl[S_i({\bf r})-S_i({\bf r}-{\bf \Delta}_i) \bigr].
$$
The second line of 
Eq.~\ref{trans2} can now be rearranged to read
$$
\times ~ \exp \Bigl[{-2\pi i \sum_{\bf r} T_{\bf r} \psi_{\bf r}
-i \sum_{\bf r}\theta_{\bf r}[ T_{\bf r} - {\bf \nabla} \cdot {\bf S}({\bf r})]
}\Bigr],
$$
from which it is now clear that the angular integration yields, up to unimportant
factors of $2\pi$,
$\prod_{\bf r} \delta_{T_{\bf r}, {\bf \nabla} \cdot {\bf S}({\bf r})}$.
Substituting this into Eq.~\ref{trans2}, performing the remaining continuous
integrals and using the $\delta$ functions to eliminate the
$T$ sum, we arrive at the remarkably simple form (ignoring unimportant prefactors)
${\cal Z}_{VM}=\sum_{\bf S}e^{-{\cal H}_{VM}}$, with
\begin{equation}
{\cal H}_{\rm VM} = \frac 1{2K}\sum_{\bf r}  |{\bf S} ({\bf r})|^2 + 
\frac 1{2h}\sum_{\bf r}\bigl ({\bf \nabla} \cdot {\bf  S} ({\bf r})\bigr)^2.
\label{ham4} 
\end{equation}

\begin{figure}[tb]
\protect \centerline{\epsfxsize=2.5in \epsfbox {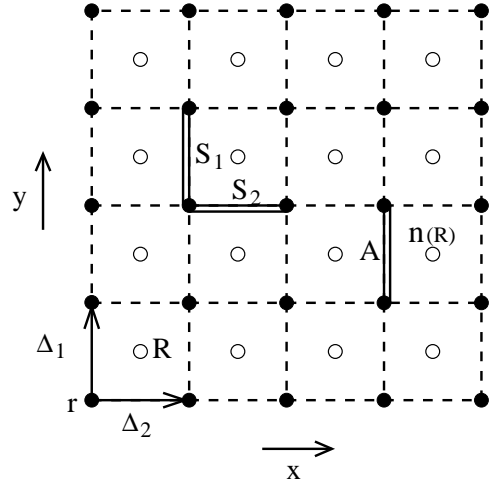}} \vskip .5cm 
\protect \caption{Real lattice sites (${\bf r}$) (black filled circles) and dual lattice sites 
(${\bf R}$) (open circles) of a square lattice. ${\bf \Delta_1}=a_0{\bf {\hat y}}$ and 
${\bf \Delta_2}=a_0{\bf {\hat x}}$ are the vectors denoting nearest neighbor bonds. $S_1$ and $S_2$ are bond integer fields that
live on the nearest neighbor bonds as shown. $S_1$ and $S_2$ can be written in terms of integers $ m_1$ and $m_2$ on the dual
sites $\lbrace {\bf R} \rbrace$ as defined in Eq.~\ref{mmap}. A 
further transformation (Eq.~\ref{hAmap})
defines two integer fields $n$ and $A$, where $A$
lives on the vertical bonds and $n$ resides on the dual lattice sites. }  
\protect \label{sqfig}
\end{figure}

In the limit $h \rightarrow 0$, the configurations entering the partition sum
must satisfy ${\bf \nabla} \cdot {\bf  S} ({\bf r}) = 0$, which implies~\cite{kadanoff} one
may write ${\bf S}$ as a two dimensional curl of an integer function $n$, 
${\bf S}({\bf r}) = {\bf {\hat z}} \times {\bf \nabla} n({\bf R}) $, where $n$ should be understood as
residing on sites of the dual lattice, ${\bf R}$ (see Fig.~\ref{sqfig}). Thus we get an
effective Hamiltonian for the $h \rightarrow 0$ limit
\begin{equation}
{\cal H}(h \rightarrow 0)_{\rm VM}
\equiv {\cal H}_{\rm DG}
= \frac 1{2K}\sum_{\bf R}  |{\bf \nabla} n({\bf R})|^2 .
\label{hdg}
\end{equation}
${\cal H}_{\rm DG}$ is the ``discrete Gaussian model'', which is one
of the simplest solid-on-solid models of an interface~\cite{weeks}.
This is well-known to be dual to the two dimensional Coulomb gas,
and so undergoes a Kosterlitz-Thouless transition; in the context of a
solid-on-solid model, it is a roughening transition~\cite{chaikin}.
It is helpful to recognize what is going on in the roughening
transition in terms of the original ${\bf S}$ variables before returning to
the case $h >0$.  Since $n({\bf R})$ represents the height of
fluctuating columns in an interface, we recognize that bonds for which
${\bf S}({\bf r}) = {\bf {\hat z}} \times {\bf \nabla} n({\bf R})\ne 0 $ actually represent
domain walls between regions of different heights.  Note that the
condition ${\bf \nabla} \cdot {\bf  S} ({\bf r}) = 0$
guarantees that the domain walls must be closed.
The important point
is that the roughening transition represents the temperature
above which such closed domain walls proliferate, i.e., 
percolate through the system.  When expressed in terms of the
Coulomb gas (i.e., vortex) model, the high temperature, rough phase
corresponds to the low temperature, bound pair phase.  

Returning to the $h>0$ phase, we see that Eq.~\ref{ham4} allows us
to work directly with the domain wall degrees of freedom.  The fact
that we allow configurations with ${\bf \nabla} \cdot {\bf  S} ({\bf r})\ne 0$
means that we are allowing open as well as closed domain walls.
In fact, locations for which ${\bf \nabla} \cdot {\bf  S} ({\bf r})\ne 0$
locate points where a domain wall comes to an end.  This is precisely
the situation one has when an interface contains {\it screw dislocations}:
locations for  ${\bf \nabla} \cdot {\bf  S} ({\bf r})\ne 0$ define the
centers of these screw dislocations, and from the second term of
${\cal H}_{\rm VM}$ we identify the core energy of the dislocation
as $1/2h$.  Remarkably, we have arrived at a theory that involves
string-like objects  (the domain walls) and vortex-like objects
(the screw dislocations), which is very much in line with what we
expected from the considerations discussed in the Introduction.
We will see that Eq.~\ref{ham4} is essentially the dual of a model
in which the string and vortex degrees of freedom are explicit;
however, we will need to introduce the vortex core energy before
this is apparent.

The above discussion shows that in our model (and its dual) the dislocations 
(vortices) are  endpoints of domain walls (strings), so that when they
are in a bound state, basic excitations of the system contain closed loops.
We would like a representation of the model that captures this physics,
presumably one that is closely related to Eq.~\ref{hdg}.  Toward this
end we need to represent the domain wall degrees of freedom ${\bf S}$
as differences.  This is most easily developed for the square lattice,
so for the rest of this section and the next few, we specialize to this case;
the triangular lattice will be dealt with in Section~\ref{TrLatt}.
We start by writing
\begin{eqnarray}
 S_1({\bf  r})&=& m_1({\bf R}={\bf r} + {\bf \Delta_1}/2-{\bf \Delta_2}/2)\cr
&-& m_1({\bf R}={\bf r} + {\bf \Delta_1}/2+{\bf \Delta_2}/2),\cr
 S_2({\bf  r})&=&m_2({\bf R}={\bf r} + {\bf \Delta_1}/2+{\bf \Delta_2}/2)\cr
&-& m_2({\bf R}={\bf r} - {\bf \Delta_1}/2+{\bf \Delta_2}/2).
\label{mmap}
\end{eqnarray}
Note that we have located the integer fields $m_1$, $m_2$ on the dual 
lattice sites $\lbrace {\bf R}\rbrace$ of the original square lattice.
In Sections~\ref{SqLatt} and~\ref{TrLatt}, we will see that representations such
as Eqs.~\ref{mmap} are convenient
for finding fixed points that respect the lattice symmetry.  For now,
we make the further
transformation
\begin{eqnarray}
m_2(X,Y)&=&n(X,Y), \cr
m_1(X,Y)&=&n(X,Y)+\sum_{x'=x_0}^{X-\Delta_2/2} A(x',Y),
\label{hAmap}
\end{eqnarray}
where ${\bf R} = (X,Y)$ are sites on the dual square lattice, $n$ and $A$
are integer fields, and the sum in the second of Eqs.~\ref{hAmap}
is along rows in the square array, starting at some reference 
line $x_0$ which can conveniently be chosen as the boundary
of the system.  If one fixes the values of $n$ and $A$ along
this boundary (e.g., $n(x_0,y)=0$, $A(x_0,Y)=0$) it is not hard
to see that there is a one-to-one invertible mapping between
the fields $(S_1,S_2)$ and $(n,A)$, so the new degrees of freedom
capture all the domain wall configurations without 
overcounting them~\cite{com2}.  
Notice in this representation it is natural to think of the
$n$ field as residing on the dual lattice sites, while the
$A$ field resides on the vertical bonds of the original lattice. This is shown in Fig.~\ref{sqfig}.

\begin{figure}[tb]
\protect \centerline{\epsfxsize=2in \epsfbox {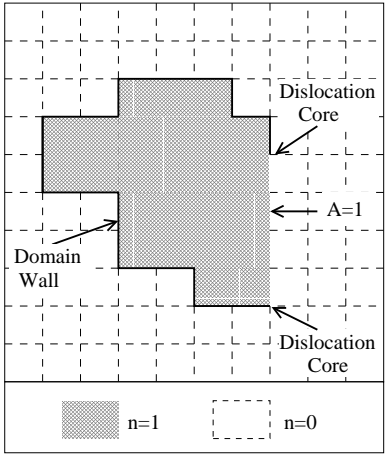}} \vskip .6cm  
\vskip .5cm
\protect \caption{A low energy configuration involving $A \ne 0$. A region of $n=1$ 
(hatched squares) is embedded in a surrounding $n=0$ region (white squares); heavy 
line represents a domain wall.  The line segment with $A=1$ cancels the domain wall 
energy for part of its length, leaving an open domain wall.}
\protect \label{fig2PRL}
\end{figure}

We now have the representation
\begin{eqnarray}
{\cal H}_{\rm VM} &=& {1 \over {2K}}
\sum_{\bf r}\bigl|({\bf \nabla}n)_{\bf r} + A(x,y+\Delta_1/2){\bf \hat{x}}\bigr|^2 \cr
&+& { 1 \over {2h}}
\sum_{\bf r}\biggl[\biggl({{\partial A} \over {\partial y}} \biggr)_{\bf r}\biggr]^2.
\label{ham_hA}
\end{eqnarray}
In this representation, it is clear that the $n$ field allows us to represent configurations
with closed domain walls. The $A$ field introduces the open domain wall configurations
in two ways: it allows us to directly occupy the vertical bonds with non-zero values,
and it allows us to remove the domain wall energy along vertical bonds of closed
domain walls -- which allows open configurations with horizontal domain
wall segments. Fig.~\ref{fig2PRL} illustrates an example of this. 

\subsection{Vortex Representation}
\label{vortex_rep}
To explicitly show the duality between dislocations and vortices,
as well as to introduce the core energy of the vortices, we need 
a representation in which the vortex degrees of freedom are explicit.
This is accomplished by eliminating $n$ using the Poisson resummation
formula, just as in the case $h=0$.
The partition sum becomes
\begin{eqnarray}
{\cal Z}_{\rm VM} &=& \sum_{\lbrace n \rbrace}\sum_{ \lbrace A \rbrace} 
\exp \Bigl[{-{\cal H}_{\rm VM}[\lbrace n \rbrace,\lbrace A \rbrace]}\Bigr], \cr
&=&  \sum_{ \lbrace A \rbrace} \sum_{\lbrace m \rbrace}\int {\cal D}\phi \; 
\exp \Bigl[ -{\cal H}_{\rm VM}[\lbrace \phi \rbrace,\lbrace A \rbrace]\cr
&-&2\pi i\sum_{\bf R} \phi({\bf R})m({\bf R})  \Bigr]. 
\label{dt1}
\end{eqnarray}
The functional integral in Eq.~\ref{dt1} may be carried through if we
represent the fields in terms of their Fourier transforms.  Writing
$\phi({\bf q}) = \sum_{\bf R} e^{i{\bf q} \cdot {\bf R}} \phi({\bf R})$,
with $N$ the number of lattice sites, and similar definitions for
$m({\bf q})$, $n({\bf q})$, and $A({\bf q})$, we have
\begin{eqnarray}
{\cal Z}_{\rm VM} \:\:\:\:\:\:\:\:\:\:\:\:\:\:\:\:\:\:\:\:\:\:\:\:\:\:\:\:\:\:\:&& \cr
=\sum_{ \lbrace A \rbrace}
 \sum_{\lbrace m \rbrace}\int {\cal D}\phi 
\exp \Biggl[ &-& {1 \over {2KN}}\sum_{\bf q}
\bigl|{\bf Q}\phi({\bf q}) + A({\bf q}){\bf \hat{x}}\bigr|^2 \cr
&-& { 1 \over {2hN}}\sum_{\bf q}|Q_y A({\bf q})|^2 \cr
&-& {{2\pi i} \over {N}} \sum_{\bf q} \phi({\bf q})m({\bf q})
 \Biggr],
\end{eqnarray}
where ${\bf Q}=(Q_x,Q_y)$, and $Q_{x}=1-e^{-i{\bf q}\cdot \Delta_2}$,
$Q_{y}=1-e^{-i{\bf q}\cdot \Delta_1}$.  Integrating out $\phi$ gives,
up to an unimportant prefactor,
\begin{eqnarray}
{\cal Z}_{\rm VM}\:\:\:\:\:\:\:\:\:\:\:\:\:\: \:\:\:\:&& \cr
= \sum_{ \lbrace A \rbrace}
 \sum_{\lbrace m \rbrace}
\exp \biggl\lbrace
&-& { 1 \over N}\sum_{\bf q} \biggl[ {1 \over {2K|Q|^2}}+{1\over {2h}} \biggr]|Q_y^2 A({\bf q})|^2 \cr
&-&{{2\pi^2 K} \over N} \sum_{\bf q}{{|m({\bf q})|^2} \over{ |Q|^2}} \cr
&+&{{2\pi i} \over N} \sum_{\bf q}{{Q_x} \over{ |Q|^2}}A({\bf q})m(-{\bf q}) \biggr\rbrace .
\label{hvq}
\end{eqnarray}
Eq.~\ref{hvq} may be conveniently represented in real space in terms of
$G_0({\bf r})={1 \over N} \sum_{\bf q} e^{-i {\bf q} \cdot {\bf r}}/ |Q|^2$
and $G_1({\bf r})={1 \over N} \sum_{\bf q} e^{-i {\bf q} \cdot {\bf r}}Q_x/Q_y |Q|^2$.
For large $r$, it is useful to note that $G_0({\bf r}) \sim \log r$
and $G_1({\bf r}) \sim \arctan (y/x)$.  We can now write
\begin{eqnarray}
{\cal Z}_{\rm VM}\:\:\:\:\:\:\:\:\:\:\:\:\:\: \:\:\:\:&& \cr
=\sum_{ \lbrace A \rbrace}
 \sum_{\lbrace m \rbrace}
\exp \biggl\lbrace
&-&2\pi^2 K \sum_{{\bf R},{\bf R'}} m({\bf R}) G_0({\bf R}-{\bf R'}) m({\bf R'}) \cr
&-& {1 \over {2K}} \sum_{{\bf r},{\bf r'}}\biggl( {{\partial A} \over {\partial y}} \biggr)_{\bf r}
G_0({\bf r}-{\bf r'}) \biggl( {{\partial A} \over {\partial y}} \biggr)_{\bf r'}
\cr
&+&2\pi i \sum_{{\bf r},{\bf R'}}\biggl( {{\partial A} \over {\partial y}} \biggr)_{\bf r}
G_1({\bf r}-{\bf r'}) m({\bf R'}) \cr
&-& {1 \over {2h}} \sum_{\bf r}
\Biggl| \biggl( {{\partial A} \over {\partial y}} \biggr)_{\bf r}\Biggr|^2
 \biggr\rbrace .
\label{hvr}
\end{eqnarray}
The logarithmic interaction among the $m$ variables establishes the fact that
they are the vortex degrees of freedom.  In this representation it is appropriate
to introduce the core energy for the vortices, after which the resulting 
model is identical to the one analyzed in Refs.~\onlinecite{nelson_book}
and~\onlinecite{jose}.
The resulting system can be considered as a {\it generalized} Villain model,
for which ${\cal Z}_{\rm GVM}=\sum_{ \lbrace m,m' \rbrace}
 e^{-{\cal H}_{\rm GVM}}$, with
\begin{eqnarray}
{\cal H}_{\rm GVM} & =&
2\pi^2 K \sum_{{\bf R},{\bf R'}} m({\bf R}) G_0({\bf R}-{\bf R'}) m({\bf R'})\cr
&+&{1 \over {2K}} \sum_{{\bf r},{\bf r'}}m'({\bf r}) G_0({\bf r}-{\bf r'})m'({\bf r'})
\cr
&-&2\pi i \sum_{{\bf r},{\bf R'}}m'({\bf r})
G_1({\bf r}-{\bf r'}) m({\bf R'}) \cr
&+& {1 \over {2h}} \sum_{\bf r} m'({\bf r})^2
 +E_c \sum_{\bf R} m({\bf R})^2.
\label{hgvm}
\end{eqnarray}
In Eq.~\ref{hgvm} we have replaced $(\partial A/\partial y)_{\bf r} \rightarrow m'({\bf r})$
to make the duality of the model apparent: upon interchanging $E_c \leftrightarrow 1/2h$
and changing $K \rightarrow 1/4\pi^2 K$, the partition sum is unchanged.  
It immediately follows that if there is an unbinding transition for the vortices
(the $m$ variables), there must also be such a transition for the dislocations 
(the $m'$ variables) at an appropriate location in the phase diagram.  This duality
was exploited in Ref.~\onlinecite{jose} to find the RG flow equations for large
$E_c$ and small $h$.  In the limit analyzed below, large $E_c$ and small $1/2h$,
we shall see that there is a dislocation unbinding transition.  The duality 
immediately tells us there must also be vortex unbinding for 
small $E_c$ and large $1/2h$.

While ${\cal H}_{\rm GVM}$ displays an elegant symmetry, it is hard to work with
directly.  The appearance of an imaginary term in the Hamiltonian makes
the configurational weights in the partition sum complex and can lead to subtle
complications, particularly if one wishes to replace the integer degrees of freedom
with continuous ones, which we will need to do to perform an RG
analysis.  Furthermore, as discussed above the physics of vortex (dislocation)
unbinding is profoundly affected by fluctuations of the strings (domain walls), which
are now only implicit in Eq.~\ref{hgvm}.   To see more clearly what is going on,
it is worthwhile to return to a representation such as Eq.~\ref{ham_hA}.
We will proceed to do this, but in doing so we will also replace  ${\cal H}_{\rm GVM}$
by a continuum model that is amenable to an RG analysis.

\subsection{Continuum Model}
\label{con-model}
To return from the vortex representation to the domain wall-dislocation representation,
one only needs to apply the Poisson resummation formula to 
the $m$ sum in 
${\cal Z}_{\rm GVM}$.  However, for the RG analysis of this system
we will want to develop an equivalent model that contains continuous rather
than integer fields, and it is at this point that it is convenient to begin doing
so.  Whereas 
\begin{eqnarray}
&&{\cal Z}_{\rm GVM}\:\:\:\:\:\:\:\cr
&=&\sum_{\lbrace n \rbrace}\sum_{\lbrace A \rbrace}\int {\cal D}\phi \;
\exp\Biggl\lbrace -{\cal H}_{\rm VM}\Bigl[\lbrace n\rightarrow \phi \rbrace, \lbrace m' \rightarrow 
{\partial A \over \partial y} \rbrace \Bigr] \cr
&-& E_c\sum_{\bf R}n^2({\bf R})
-2\pi i\sum_{\bf R}\phi({\bf R})n({\bf R}) \Biggr\rbrace,
\cr
&\equiv& \sum_{\lbrace A \rbrace}\int {\cal D}\phi \prod_{\bf R}
\sum_{n({\bf R})=-\infty}^{\infty} 
\exp \Biggl\lbrace-{\cal H}_{\rm VM}\Bigl[ \phi,   
{\partial A \over \partial y}\Bigr] \cr
&-&  E_c\sum_{\bf R}n^2({\bf R})
 - 2\pi i\sum_{\bf R}\phi({\bf R})n({\bf R}) \Biggr\rbrace,
\nonumber
\end{eqnarray}
is an exact representation of the generalized Villain model, we truncate the
sums $\sum_{n({\bf R})=-\infty}^{\infty}$ to $\sum_{n({\bf R})=-1}^{1}$.
For large $E_c$ this is an excellent approximation; vortices with large
topological charge play little role in the properties of the system when
the fugacity is small.   Using $\sum_{n=-1}^{1}\exp (-2\pi i n\phi-E_c\sum_{\bf R}n^2)=
1+y\cos(2\pi\phi) \approx \exp[y\cos(2\pi\phi)]$ with $y=2e^{-E_c}$ twice the
fugacity, we arrive at an intermediate model,
\begin{eqnarray}
{\cal Z}'=\sum_{\lbrace A \rbrace}\int {\cal D}\phi \;
\exp \Biggl\lbrace &-& {\cal H}_{\rm VM}\Bigl[ \phi,    
{\partial A \over \partial y }\Bigr] \cr
&+& y\sum_{\bf R}\cos(2\pi\phi) \Biggr\rbrace.
\label{zprime}
\end{eqnarray}
At this point we have one continuous and one integer field.  To
go over to a fully continuous model we make the replacement
${\cal Z}' \rightarrow {\cal Z} = \int {\cal D}\phi \int {\cal D}a \; e^{-{\cal H}_{\rm eff}[\phi,a]}$,
with
\begin{eqnarray}
{\cal H}_{\rm eff}=\int d^2 r \biggl[ &&{1 \over {2K}} \big|{\bf {\nabla}} \phi({\bf r}) 
+ a({\bf r}) {\bf \hat{x}} \big|^2
+ {{1} \over {2h}} \biggl( {{\partial a} \over {\partial y}} \biggr)^2 \nonumber \\
-&y&  \cos\bigl( 2\pi \phi({\bf r}) \bigr) + y \cr
- &y_a& \cos\bigl( 2\pi a({\bf r}) \bigr) +y_a \biggr] .
\label{heff}
\end{eqnarray}
In Eq.~\ref{heff}, we have gone over from a lattice to a continuum representation, taking
our lattice constant $a_0 \equiv 1$ as our unit of length, and we have subtracted
constants so that the ground state energy is zero.
Although our replacement of the integer field $A$ with the continuous field
$a$ is not a controlled approximation in the same sense as when we introduced
$\phi$ in favor of $n$, we have constructed ${\cal H}_{\rm eff}$ so that it
has the same symmetry properties under translations of $a \rightarrow a+m$, with
$m$ an integer, as had ${\cal H}_{\rm GVM}$, using the $y_a$ cosine term.  
Thus we expect  ${\cal H}_{\rm eff}$
and ${\cal H}_{\rm GVM}$
to have the same phases and types of transitions among them~\cite{cardy}.
Our replacement of ${\cal H}_{\rm GVM}$ with ${\cal H}_{\rm eff}$ is in fact
no better or worse than our initial replacement of ${\cal H}_{XY}$ with ${\cal H}_{\rm VM}$.
${\cal H}_{\rm eff}$ is the model that we will focus on in our RG analysis.

\begin{figure}[tb]
\protect \centerline{\epsfxsize=3.2in \epsfbox {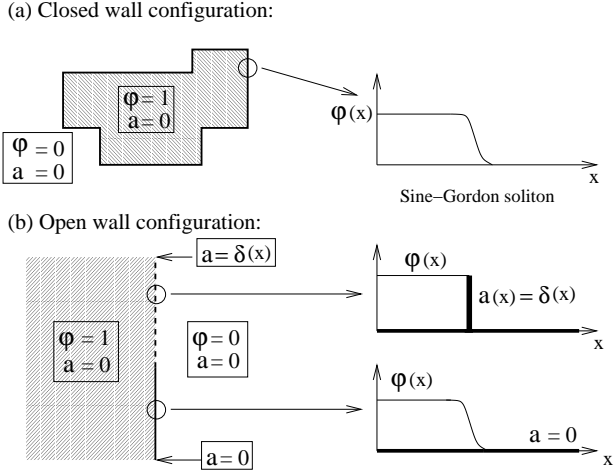}} \vskip .5cm 
\protect \caption{Different low energy soliton configurations of $(\phi, a)$ for the effective Hamiltonian
${\cal H}_{\rm eff}$ defined in Eq.~\ref{heff}. (a) For closed wall configurations with $a(x)=0$, ${\cal H}_{\rm eff}$ 
is a two-dimensional sine-Gordon equation with sine-Gordon
solitons. (b) $a(x)=\delta (x)$ for large positive $y$, $a(x)=0$ for large negative $y$, and
$\phi(x,y) \rightarrow 1$ as $x \rightarrow -\infty$, 
$\phi(x,y) \rightarrow 0$ as $x \rightarrow \infty$. The domain wall has finite energy per unit length as
$y \rightarrow -\infty$, but vanishing energy per unit length as $y \rightarrow \infty$. The resulting
configuration is a screw dislocation.}
\protect \label{domainwall}
\end{figure}

It is useful to recognize that the low-energy configurations of $(\phi,a)$ in
${\cal H}_{\rm eff}$  mirror those of the Villain model (Eq.~\ref{ham4})
with which we started.  For example, if we consider configurations with
$a=0$, then ${\cal H}_{\rm eff}$ is a two dimensional sine-Gordon
model, supporting kink excitations~\cite{raja} that are directly analogous
to domain walls.  ${\cal H}_{\rm eff}$ also supports solitons that represent
screw dislocations. We can see this by considering configurations in
which $\phi(x,y) \rightarrow 1$ as $x \rightarrow -\infty$ and 
$\phi(x,y) \rightarrow 0$ as $x \rightarrow \infty$.  For $a=0$
this will force in a domain wall running along the ${\bf \hat{y}}$ direction.
However, if we set $a(x,y)= \delta (x)$, then we can 
produce a configuration of vanishing energy and
satisfy the boundary
condition on $\phi$ with $\phi(x,y)=\Theta(-x)$, where $\Theta(-x)$ is a step
function.   A screw dislocation is forced into the system when we
impose the boundary conditions $[\phi(x,y),a(x,y)] \rightarrow 
[\phi_{\rm SGS}(x),0]$ as $y \rightarrow -\infty$, and 
$[\phi(x,y),a(x,y)] \rightarrow 
[\Theta(-x),\delta (x)]$ as  $y \rightarrow \infty$,
where $\phi_{\rm SGS}(x)$ is the kink soliton of the
sine-Gordon model~\cite{raja}.  These boundary conditions
guarantee that the domain wall must end somewhere in the
bulk of the sample.  This endpoint is the screw dislocation core. This is pictorially
shown in Fig.~\ref{domainwall}.

Finally, while most parameters entering  ${\cal H}_{\rm eff}(K, h$, and $y$)
are directly related to parameters of the generalized Villain model
($K, h$, and $E_c$), the value of $y_a$ in terms of these
parameters is not immediately obvious.  The connection can be
made by recognizing that $y_a$ plays a key role in determining the
dislocation core energy in ${\cal H}_{\rm eff}$: if $y_a$ is excessively
large, then variations of $a$ in the ${\bf \hat{y}}$ direction become
energetically very expensive, so that the core of a dislocation, where
$a$ for example varies from $0$ to $\delta (x)$, becomes high in energy.

Unfortunately the screw dislocation is a rather complicated soliton so
that it is not obvious how one might obtain an analytic expression
for its energy.  We can, however, make a rough estimate for $y_a$ by looking at
the dislocation configuration in the Hamiltonian implicit in the 
partition sum for ${\cal Z}'$ in Eq.~\ref{zprime}.  In  Eq.~\ref{zprime},
$A$ is an integer variable residing on the vertical bonds, and 
we create a dislocation near the origin by setting 
$A(x,Y)=-1$ for $x=0,~Y>0$, and $A=0$ elsewhere, 
and introducing the same
boundary conditions on $\phi$ described two paragraphs above.
It is clear for large negative $Y$, $\phi$ will contain a sine-Gordon
kink soliton, whereas for large positive $Y$, it has the form of a
step function.  The $\phi$ kink presumably narrows from the sine-Gordon 
width to zero in the vicinity of the origin.  The energy of
the configuration will take the form $\varepsilon = \varepsilon_{\phi}
+1/2h$, the first contribution representing the energy stored in the
$\phi$ field and its interaction with $A$, and the second the energy
cost for the step down in $A$ at the origin.  
In going over to ${\cal H}_{\rm eff}$, where $A$ is replaced by
the continuous field $a$,  if we assume the contribution  $\varepsilon_{\phi}$
is unchanged, then we can approximately match the dislocation
core energies by matching the cost of the step down in $a$ with
$1/2h$.  This is estimated by keeping only the second and fifth
terms of the right hand side of Eq.~\ref{heff}, so that the step is again a sine-Gordon
kink.  The energy of the kink is~\cite{raja} $8\sqrt{y_a/h}$;
we obtain the match by setting $y_a \approx 1/256h$.

With this estimate, it is possible to connect the phases
of ${\cal H}_{\rm eff}$ with those of the GVM and $XY$ models.
We now address this with an RG analysis.

\section{Renormalization Group Analysis}
\label{RGanalysis}
\subsection{Derivation of Scaling Relations}
\label{Scaling_Relation}
\begin{figure}[tb]
\protect \centerline{\epsfxsize=3in \epsfbox {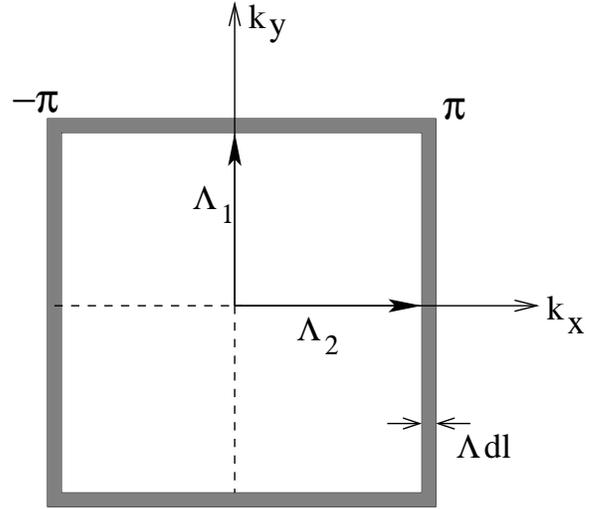}} \vskip .5cm 
\protect \caption{Momentum shell in reciprocal space for the square lattice. ${\bf \Lambda}_1$ and 
${\bf \Lambda}_2$ are 
equal length vectors of magnitude $\pi$ ($a_0=1$) to the center of the BZ edges. The width of the shaded region is $\Lambda d\ell$.} 
\protect \label{squarefig}
\end{figure}

Our analysis of  ${\cal H}_{\rm eff}$ proceeds with a momentum
shell renormalization group procedure~\cite{chaikin} that is perturbative in
$y$ and $y_a$.   This involves dividing the fields
into long and short wavelength components: $\phi({\bf r})=
\phi^{<}({\bf r})+\phi^{>}({\bf r})$, with
\begin{eqnarray}
\phi^{<}({\bf r})&=& \int_{|q_x|,|q_y|<\Lambda/b}
{{d^2{\bf q}} \over {(2\pi)^2}}\;e^{-i{\bf q} \cdot {\bf r}}\phi({\bf q}),   \cr
\phi^{>}({\bf r})&=&\biggl( \int_{\Lambda/b<|q_x|<\Lambda}
+ \int_{\Lambda/b<|q_y|<\Lambda}\biggr)
{{d^2{\bf q}} \over {(2\pi)^2}}\;e^{-i{\bf q} \cdot {\bf r}}\phi({\bf q}).
\label{momshell}
\end{eqnarray}
There is an analogous decomposition for $a$.
In Eqs.~\ref{momshell}, $\Lambda=\pi/a_0$, $b=e^{\ell}$ is the
rescaling factor, and we are including
in $\phi^{>}$  momentum components in a square shell at the edge
of the Brillouin zone, as illustrated in Fig.~\ref{squarefig}.

As we shall see, the fixed points we are interested in are not orientationally isotropic,
so it is important to choose the momentum shell so that it respects
the symmetry of the lattice. (When we address the triangular
lattice below, our momentum shell will be a hexagon.)

As usual~\cite{cardy}, the analysis proceeds by integrating out
$\phi^{>}$ and $a^{>}$ from the
partition function ${\cal Z}$.  For $y,~y_a=0$ we then have
\begin{eqnarray}
{\cal Z}(\ell) &=& e^{-F(\ell)} \int {\cal D} \phi^{<}\int {\cal D} a^{<} \cr 
\times \exp \biggl[&-& {1 \over {2K}}\int^{<}{{d^2{\bf q}} \over {(2\pi)^2}}  
\biggl[\bigl|-i{\bf q} \phi^{<}({\bf q}) + a^{<}({\bf q}){\bf \hat{x}} \bigr|^2 \cr
&+& \xi^2  |a^{<}({\bf q})|^2 \biggr] \biggr],
\label{zeroth_ord}
\end{eqnarray}
where $e^{-F(\ell)}$ is a prefactor that comes from the integrals
over $\phi^{>}$ and $a^{>}$,  $ \int^{<}$ indicates an integral
in the inner part of the Brillouin zone (i.e., inside the shaded region
in Fig.~\ref{squarefig})
and $\xi=\sqrt{K/h}$ is the width
of a string connecting a highly separated vortex-antivortex pair~\cite{girv}.
We can bring what remains of the Hamiltonian back to its original
form (for  $y,~y_a=0$)  with the scaling transformation
\begin{eqnarray}
{\bf q} = {\bf q^{\prime} } /b \quad&, &\quad {\bf r} = b {\bf r^{\prime} },\cr
 \phi^{<}({\bf r})=\phi^{\prime}({\bf r^{\prime} })~&,&~
a^{<}({\bf r}) = a^{\prime}({\bf r^{\prime} }) /b. 
\label{scaling}
\end{eqnarray}
We then have in real space
\begin{eqnarray*}
{\cal Z}(\ell) = e^{-F(\ell)}&& \int {\cal D} \phi^{\prime}\int {\cal D} a^{\prime}\;\cr
\times \exp \Biggl\lbrace &-& {1 \over {2K'}}\int d^2r'  \biggl[
\bigl| {\bf \nabla} \phi^{\prime}({\bf r'}) + a^{\prime}({\bf r'}){\bf \hat{x}} \bigr|^2 \cr
&+& \xi^{\prime 2}  |a^{\prime}({\bf r'})|^2 \biggr]^2\Biggr\rbrace,
\end{eqnarray*}
with $K'=K$, $ \xi^{\prime 2}= \xi^{ 2}/b^2$.  For a very narrow
shell ($\ell \rightarrow d\ell$, $b \rightarrow 1+d\ell$), the differential
changes in $K$ and $\xi^2$ are just
\begin{eqnarray}
{{dK} \over {d\ell}} &=& 0, \cr
{{d\xi^2} \over {d\ell}} &=& -2\xi^2.
\label{scale_rel_1}
\end{eqnarray}

Thus we have a set of fixed points whose stability with respect
to the cosine perturbations we need to assess.  For the $y\cos(2\pi \phi)$
perturbation we will proceed in the standard 
fashion~\cite{chaikin,gogolin}, and we will see 
near the fixed points of interest it has a large anomalous dimension --
the perturbation is highly irrelevant.  The $y\cos(2\pi a)$ perturbation,
on the other hand, is rather unusual.  The scaling relation for $a$ in
Eq.~\ref{scaling} causes it to shrink as one integrates out the short
wavelength fluctuations of the system.  This suggests we should
expand
\begin{equation}
y_a \cos\bigl( 2\pi a \bigr) -y_a = \sum_{n=1}^{\infty} {y_{2n} \over {(2n)!}}
(-1)^n (2\pi a)^{2n},
\nonumber
\end{equation}
with the initial values $y_{2n}(l=0) = y_a$.  Our effective Hamiltonian is now
\begin{eqnarray}
{\cal H}_{\rm eff}=\int d^2 r \biggl[&& {1 \over {2K}} \big|{\bf \nabla} \phi({\bf r}) + a({\bf r}) {\bf \hat{x}} \big|^2
+ {{1} \over {2h}} \biggl( {{\partial a} \over {\partial y}} \biggr)^2 \nonumber \\
-&y& \cos\bigl( 2\pi \phi({\bf r}) \bigr) + y \cr
&-& \sum_{n=1}^{\infty} {y_{2n} \over {(2n)!}}
(-1)^n (2\pi a)^{2n}\biggr] .
\label{heff_exp}
\end{eqnarray}
If we simply substitute $a({\bf r}) \rightarrow a^{\prime}({\bf r'})/b$ in the
last term of Eq.~\ref{heff_exp}, we see that the 
coefficients $y_{2n}\sim b^{2-2n}$.  This suggests, as we will
confirm below, that most of the terms in this expansion are irrelevant.  However,
the Gaussian (i.e., first) term is special because it neither grows
nor shrinks, suggesting we should incorporate it into our fixed
point.  Writing $\rho(\ell) \equiv 4\pi^2 y_2(\ell)$, we take our
unperturbed Hamiltonian to be
\begin{eqnarray}
{\cal H}_{0}=\int d^2 r &\biggl[& {1 \over {2K}} \big|{\bf \nabla} \phi({\bf r}) 
+ a({\bf r}) {\bf \hat{x}} \big|^2
+{1 \over 2}\rho a({\bf r})^2 \cr
&+& {{\xi^2} \over {2K}} \biggl( {{\partial a} \over {\partial y}} \biggr)^2
\biggr].
\label{h0}
\end{eqnarray}

To first order in $y, y_a$, the partition function takes the form
\begin{eqnarray}
{\cal Z}& =& \int {\cal D} \phi^{<} \int {\cal D}\phi^{>} 
\int {\cal D} a^{<} \int {\cal D}a^{>}e^{{-\cal H}_{0}}
 \cr
&\times&
\Biggl\lbrace 1-y\int d^2r \cos[2\pi \phi^{<}({\bf r})+2\pi \phi^{>}({\bf r})] \cr
 &-&\int d^2r \sum_{n=2}^{\infty} {y_{2n} \over {(2n)!}}
(-1)^n \Bigl[2\pi a^{<}({\bf r})+2\pi a^{>}({\bf r})\Bigr]^{2n} \Biggr \rbrace.
\label{first_ord}
\end{eqnarray}
The ``1'' term above has essentially been discussed already;
the only difference one gets after integrating out $\phi^{>},~a^{>}$
from Eq.~\ref{zeroth_ord} is that $F(\ell)$ now has a correction
of order $y_a$ due to the $\rho$ term appearing in ${\cal H}_0$.
Thus the scaling relations Eqs.~\ref{scale_rel_1} are correct
to first order in $y$ and $y_a$.  We next wish to integrate out
the short wavelength fields in the middle term of Eq.~\ref{first_ord}.
Calling this contribution ${\cal Z}_y$, the integral is
\begin{eqnarray}
{\cal Z}_y &\equiv& -y\int {\cal D} \phi^{<} \int {\cal D}\phi^{>} 
\int {\cal D} a^{<} \int {\cal D}a^{>}e^{{-\cal H}_{0}} \cr
&\times & \Biggl\lbrace \int d^2r \cos[2\pi \phi^{<}({\bf r})+2\pi \phi^{>}({\bf r})] \Biggr \rbrace, \cr
&=& 
 -ye^{-F(\ell)} \cr
&\times & \exp \Biggl\lbrace -{K \over 2} \int^{>}d^2{\bf q} 
\biggl[ {q^2-{{q_x^2} \over {1+K\rho+\xi^2q_y^2}}}\biggr]^{-1}\Biggr\rbrace \cr
&\times &\int {\cal D} \phi^{<}\int {\cal D} a^{<}e^{{-\cal H}_{0}^{<}}
\int d^2r \cos[2\pi \phi^{<}({\bf r})],
\end{eqnarray}
where ${\cal H}_{0}^{<}$ is obtained by  Fourier transforming
the right hand side of Eq.~\ref{h0}, and then dropping wavevectors
in the resulting $\bf q$ integral that are in the momentum shell.
We can already see the crucial role played by the parameter $\rho$: for
$\rho=0$, the $\int^{>}d^2{\bf q}$ integral is divergent, and it is apparent
that ${\cal Z}_y$ vanishes in the thermodynamic limit.  Since the values
of $\rho$ in which we are interested are indeed small, it is apparent that the $y\cos(2\pi \phi)$
is going to be strongly irrelevant.  We estimate the $\int^{>}d^2{\bf q}$ integral 
by evaluating it for $\xi=0$ since this variable is irrelevant (Eq.~\ref{scale_rel_1}).
For small $\rho$, one finds
$$
 \int^{>}d^2{\bf q}\; 
\biggl[ {q^2-{{q_x^2} \over {1+K\rho}}}\biggr]^{-1}=
\pi K\sqrt{{1 + K\rho} \over {K\rho}} \log b +{\cal O}(\rho).
$$
Rescaling the fields and lengths, we can return ${\cal Z}_y$ to its original form, writing
$$
{\cal Z}_y= -y'e^{-F(\ell)}
\int {\cal D} \phi \int {\cal D} a \; e^{{-\cal H}_{0} }
\int d^2r \cos[2\pi \phi ({\bf r})],
$$
with
$$
y'=yb^{2-\pi K\sqrt{{1 + K\rho} \over {K\rho}}},
$$ 
where again we are focusing on small $\rho$.  Recalling $b=1+d\ell + {\cal O}(d\ell^2)$,
we now have the scaling relation for $y$,
\begin{equation}
{{dy} \over {d\ell}} = \Biggl[2-\pi K\sqrt{{1 + K\rho} \over {K\rho}}\Biggr]y.
\label{scale_rel_2}
\end{equation}

Because $\rho$ is small (recall its initial value is ${\cal O}(y_a)$, and we shall
see it shrinks under renormalization), for the parameters of interest $y$ shrinks
very rapidly as we integrate out short wavelengths.  The $y\cos (2\pi \phi)$ term
is thus strongly irrelevant, suggesting that the system is in a ``rough'' phase -- domain
walls are proliferated through the system.  We will discuss the implications of
this below.

\begin{figure}[tb]
\protect \centerline{\epsfxsize=3in \epsfbox {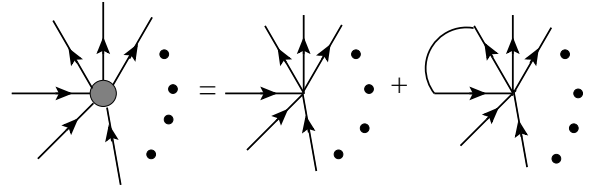}} \vskip .5cm 
\protect \caption{Pictorial representation of Eq.~\ref{renorm}. Integral over short wavelength fields [$\phi^> ({\bf r})$] of the $y_{2n}$ term introduces a renormalization of the
$y_{2n-2}$ term. The graphical decomposition of a $2n$ point vertex is shown. The first diagram on the right 
hand side is a
tree diagram that represents all the long-wavelength fields [$\phi^< ({\bf r})$] which correspond to the first
term in Eq.~\ref{renorm}. Contraction between a pair of legs is shown in the second diagram and it represents the
second term in Eq.~\ref{renorm}. The black dots in the figure represent other legs. } 
\protect \label{vertex}
\end{figure}

We are left finally with deriving the scaling relations for the polynomial terms in $a$.
These are also handled in a standard fashion~\cite{chaikin}.   We first consider
a term in the last sum of Eq.~\ref{first_ord}.  Expressed in terms of the Fourier
components of the fields, such a term may be written as
\begin{eqnarray}
{\cal Z}_{y_{2n}} &\equiv&
- \int {\cal D} \phi^{<} \int {\cal D}\phi^{>} 
\int {\cal D} a^{<} \int {\cal D}a^{>}e^{{-\cal H}_{0}}\cr
&\times&{y_{2n} \over {(2n)!}}(-1)^n \int  {{d^2 {\bf q_1}}}{{d^2 {\bf q_2}} }
\dots {{d^2 {\bf q_{2n}}}} \cr
&\times & (2\pi)^2 \delta({\bf q_1}+ {\bf q_2}+ \dots + {\bf q_{2n}} ) \cr
&\times & a({\bf q_1})a({\bf q_2})\dots a({\bf q_{2n}}).
\label{zy2n}
\end{eqnarray}
In writing this in terms of short and long-wavelength contributions, we wish to
separate each of the $a({\bf q_i})$'s in the product above by the location of 
${\bf q_i}$ in the Brillouin zone.  One thus has
\begin{eqnarray}
 &&a({\bf q_1})a({\bf q_2})\dots a({\bf q_{2n}}) 
 \rightarrow \cr 
 && \cr
 && a^{<}({\bf q_1})a^{<}({\bf q_2})\dots a^{<}({\bf q_{2n}}) \cr
&+&
{{2n(2n-1)} \over 2} a^{>}({\bf q_1})a^{>}({\bf q_2})
a^{<}({\bf q_3})\dots a^{<}({\bf q_{2n}}) +...,
\label{a_prod}
\end{eqnarray}
where we have dropped the term with just one  $a^{>}$, anticipating
this will integrate to zero in the functional integral.  Because our
momentum shell has width $d\ell$, we need only retain the two
terms explicitly shown in Eq.~\ref{a_prod}; terms with more 
factors of $a^{>}$ involve higher powers of $d\ell$ and vanish
when we take $d\ell \rightarrow 0$.  Substituting this expansion into
Eq.~\ref{zy2n}, we obtain
\begin{eqnarray}
{\cal Z}_{y_{2n}} &=&
-y_{2n} \int {\cal D} \phi^{<}\int {\cal D} a^{<}
e^{-{\cal H}_0^{<}}\cr
&\times&
e^{-F(\ell)}\int d^2r
\Biggl\lbrace {{(-1)^n} \over {(2n)!}}
\Bigl[2\pi a^{<}({\bf r})\Bigr]^{2n} \cr
&-&{{(-1)^{n-1}} \over {(2n-2)!}} 2\pi^2 {\cal L}(\rho,\xi) d\ell
 \Bigl[2\pi a^{<}({\bf r})\Bigr]^{2n-2}\Biggr\rbrace,
\label{renorm}
\end{eqnarray}
with
\begin{eqnarray}
e^{-F(\ell)}{\cal L}(\rho,\xi) d\ell &=& 
\int {\cal D} \phi^{>}\int {\cal D} a^{>}
e^{-{\cal H}_0^{>}} \cr
&\times & \int^{>}{{d^2{\bf q}} \over {(2\pi)^2}} \; a^{>}(-{\bf q})a^{>}({\bf q}).
\label{shell_int}
\end{eqnarray}
Eq.~\ref{renorm} shows that the integral over short wavelengths
of the $y_{2n}$ term introduces a renormalization of the $y_{2n-2}$ term.
This is illustrated graphically in Fig.~\ref{vertex}.
Rescaling fields and lengths, we see that ${\cal Z}_{y_{2n}}$ may also be returned
to its original form,
\begin{equation}
{\cal Z}_{y_{2n}} = -y_{2n}' e^{-F(\ell)} \int {\cal D} \phi\int {\cal D} a \;
e^{-{\cal H}_0}\biggl\lbrace {{(-1)^n} \over {(2n)!}}
\Bigl[2\pi a({\bf r})\Bigr]^{2n}\biggr\rbrace,
\nonumber
\end{equation}
with
$$
y_{2n}' = y_{2n}b^{2-2n} - 2\pi^2 [{\cal L}(\rho,\xi) d\ell] ~ y_{2n+2}.
$$
This allows us to write down our last scaling relation,
\begin{equation}
{{dy_{2n}} \over {dl} } = -(2n-2)y_{2n} -2\pi^2 {\cal L} (\rho,\xi) y_{2n+2}.
\label{scale_rel_3}
\end{equation}
A slightly subtle but important question arises when we recall
$ 4\pi^2 y_2 =\rho$.  Should Eq.~\ref{scale_rel_3} be applied
to $\rho$, even though it appears in ${\cal H}_0$ and is not
included in the ${\cal Z}_{y_{2n}}$'s?  The answer is yes;
the integration of the $a^4$ term does produce a Gaussian
term that renormalizes the ${1 \over 2} \rho a^2$ in ${\cal H}_0$ .
This is easily seen if we reexponentiate the ${\cal O}(y_a)$ contribution 
to the partition function.  The renormalizations of the $y_{2n}$ terms
follow through exactly as we have described above, and one can then
see explicitly that the $y_4$ term renormalizes $\rho$.  Thus we 
can write for the $n=1$ case of Eq.~\ref{scale_rel_3}
\begin{equation}
{{d\rho} \over {dl} } = -8\pi^4 {\cal L} (\rho,\xi) y_{4}.
\label{scale_rel_4}
\end{equation}

The derivation of the scaling relations is completed by writing
down an explicit expression for ${\cal L} (\rho,\xi) $.  
The functional integrals in Eq.~\ref{shell_int} yield
\begin{equation}
{\cal L} (\rho,\xi) d\ell =  \int^{>} {{d^2{\bf q}} \over {(2\pi)^2}} 
{{2K q^2} \over {q_y^2 +[K\rho + \xi^2q_y^2]q^2}}.
\label{int_line}
\end{equation}
This integral can be computed exactly in a tedious but in 
principle straightforward calculation.  The result is displayed
in Appendix A.  For small $\rho$, the result (Eq.~\ref{int_line_exact})
may be expanded to yield
\begin{equation}
{\cal L} \approx {{K \Lambda^2} \over
{\pi \sqrt{K\rho(1+\xi^2\Lambda^2)}}}.
\label{int_line_approx}
\end{equation}
Eq.~\ref{int_line_approx} once again demonstrates the importance of
the parameter $\rho$: as $\rho \rightarrow 0$, ${\cal L}$ diverges,
suggesting that the ${1 \over 2} \rho a^2$ term in ${\cal H}_0$
is irrelevant if the initial value of $\rho$ is small enough.  We will
see this is indeed the case.

Eqs.~\ref{scale_rel_1},~\ref{scale_rel_2},~\ref{scale_rel_3}, and \ref{scale_rel_4}
are the scaling equations for our problem.  Before presenting results
obtained from numerically integrating them in Sec.~\ref{discussion}, we discuss the fixed 
points of the equations and their interpretation.

\subsection{Fixed Points: Bound and Unbound Dislocation States}
\label{bound_state}
It is easy to see that fixed points of Eqs.~\ref{scale_rel_1},~\ref{scale_rel_2},
~\ref{scale_rel_3}, and \ref{scale_rel_4} occur for $y=0$, $y_{2n \ge 4}=0$,
and $\xi^2=0$, so that the generic fixed point Hamiltonian has the form
\begin{equation}
H_{*}=
\int d^2 r \biggl[ {1 \over {2K}} \big|{\bf \nabla} \phi({\bf r}) + a({\bf r}) {\bf \hat{x}} \big|^2
+ {1 \over 2} \rho   a({\bf r})^2
\biggr] .
\label{FP}
\end{equation}
We will see in our numerical integration of the scaling relations that
any (small) value of $\rho$ is an attractive fixed point.  This will
give us an unbroken line of fixed points.  In spite of the $\rho=0$ point
being connected to the $\rho>0$ fixed points, these two cases 
are physically different.  The reason goes back to the role $a$ plays
in allowing us to form separated dislocation pairs.   
We demonstrated in the last section that one can remove the energy of
a section of domain wall in the $\phi$ field when $\rho=0$,  by creating a narrow
line segment where $a \ne 0$ along some part of the domain wall (see
Fig.~\ref{domainwall}).  Because the $y\cos(2\pi\phi)$ term is irrelevant, we
understand that domain walls in $\phi$ proliferate through the system.
Thus, if we consider only $a=0$ configurations, 
we can find arbitrarily large domain wall  loops.  
Such loops can be broken open 
to create dislocation pairs.  The only extra energy cost for doing this
to a closed loop of length $L$
is twice the dislocation core energy;  the loop can be deformed to move
the dislocations a distance $\sim L$ apart, without any further cost in energy.
This finite core energy is more than balanced by the entropy associated
with the number of open domain wall configurations for large $L$, so 
that large open domain wall configurations -- highly separated dislocation
states -- are not suppressed by the dislocation core energy.  We thus identify
the $\rho=0$ fixed point as an unbound dislocation state.

For any non-vanishing value of $\rho$, the situation for large enough $L$ becomes
profoundly different.  The ``mass'' term for $a$ tells us that we cannot 
completely eliminate the domain wall energy.  If we create a dislocation
pair by deforming a closed loop, there is a residual attractive interaction
between them.  {\it We can thus identify $\rho$ as a renormalized
string tension for the dislocation pair.  When $\rho>0$, the dislocations
are bound into pairs.}   It is interesting to note that
because the $a$ field lives only on the vertical bonds of the underlying lattice,
we do not get the configurational entropy that the domain walls in
the $\phi$ field has -- so even though the residual ``string'' connecting
the dislocations may have much lower {\it energy} per unit length than 
the closed domain wall loops, the {\it free} energy of closed domain walls
can vanish, while that of the residual strings remains finite.

While it is tempting to interpret the energy of the residual strings as leading
to linear confinement, this is not correct because it ignores the
effect of fluctuations in $\phi$.  A simple interpretation may be had
if we interpret $\rho>0$ as telling us that the``integerness" of the
original $A$ field cannot be ignored -- the $\cos(2\pi a)$ term 
contains physics which is relevant at long wavelengths --
even while the irrelevance
of $y$ tells us we may treat $\phi$ as a continuous field and ignore
the $\cos(2\pi\phi)$ perturbation.  This suggests we can return to
${\cal{Z}'}$  (Eq.~\ref{zprime}),  setting $y=0$, which allows
us to integrate out $\phi$.  Up to unimportant prefactors, one finds
${\cal{Z}'} \propto \sum_{\lbrace A \rbrace}e^{-{\cal H}_{CG}}$,
with
\begin{eqnarray*}
{\cal H}_{CG} &=&{1 \over {2K}} \sum_{{\bf r},{\bf r'}}
\biggl( {{\partial A} \over {\partial y}} \biggr)_{\bf r}
G_0({\bf r}-{\bf r'}) \biggl( {{\partial A} \over {\partial y}} \biggr)_{\bf r'} \cr
&+& {1 \over {2h}} \sum_{\bf r}
\Biggl| \biggl( {{\partial A} \over {\partial y}} \biggr)_{\bf r}\Biggr|^2
 \biggr\rbrace.
\end{eqnarray*}
Recalling that $({\partial A}/{\partial y})$ is a representation
of the dislocation field, we see that we have come to a Coulomb gas
Hamiltonian!   This suggests we should think of the $\rho>0$ state
as a set of logarithmically interacting dislocations, rather than
linearly confined ones.  However, in coming to $H_{CG}$ 
we understood $\rho>0$ as signaling that we cannot ignore the
discreteness of the underlying charge.  This is the usual situation
when the charges are in a bound state.  If one can treat 
$({\partial A}/ {\partial y}) \rightarrow ({\partial a}/{\partial y})$ 
as a continuous field, the state may be interpreted as a ``charged liquid''~\cite{getref}
and the dislocations may be treated as unbound.  This is only valid if $\rho=0$.

An alternative look at the fixed points may be had by reexpressing them in
terms of the vortices.  To do this, we return $\phi$ to its integer form $n$
but keep $a$ as a continuous field, adopting the fixed point form for
its energetics.  Applying the Poisson resummation formula to $n$ brings
us to an effective partition function $\tilde{\cal Z}'=\sum_{\lbrace m \rbrace} \int{\cal D} a \;
e^{-\tilde{\cal H}'}$ with
\begin{eqnarray*}
\tilde{\cal H}'=
{1 \over N} \sum_{\bf q} &\biggl\lbrace& \biggl[ {1 \over {2K|Q|^2}} \biggr]|Q_y^2 a({\bf q})|^2
+{{\rho} \over 2} |a({\bf q})|^2 \cr
&+& 2\pi^2 K{{|m({\bf q})|^2} \over{ |Q|^2}}
-2\pi i{{Q_x} \over{ |Q|^2}}a({\bf q})m(-{\bf q}) \biggr\rbrace.
\end{eqnarray*}
Integrating out $a$ gives, again up to unimportant prefactors, 
$\tilde{\cal Z}'=\sum_{\lbrace m \rbrace}
e^{-\tilde{\cal H}}$, with the reduced Hamiltonian $\tilde{\cal H}$
now
$$
\tilde{\cal H}=
{{2\pi^2 K} \over N} \sum_{\bf q}{{|m({\bf q})|^2} \over{ |Q|^2}}
\Biggl[ 1+{{|Q_x|^2} \over {|Q_y|^2+\rho K  |Q|^2}} \Biggr].
$$
The energetics of the vortices is profoundly different for 
$\rho>0 $ and $\rho=0$.  This is most clearly seen by considering a configuration
for a single vortex at the origin, $m({\bf R})=\delta_{{\bf R},0}$, in a finite
size system of linear dimension $L$.
The energy of such a configuration has the form
\begin{eqnarray}
<\tilde{\cal H}>_{1v}&\sim&(1+C)\log L, \quad\quad (\rho>0) \cr
&\sim& L,  \quad\quad (\rho = 0)
\label{1v}
\end{eqnarray}
where $C=\int_0^{2\pi} d\theta\; [{\cos^2 \theta}/ (\sin^2\theta +\rho K)]$.
We find the usual logarithmic energy for a vortex when $\rho>0$, although the
correction $C$ to the coupling constant becomes very large for small $\rho$.
The effect of the dislocations thus is a strong {\it upward} renormalization
of the effective $K$ in this state.  For $\rho=0$, by contrast, the energy of a single vortex
configuration grows linearly with $L$.  Thus, the state that we identified as the
unbound dislocation state becomes, in the dual representation, a linearly confined vortex
state.  By contrast, the $\rho>0$ state represents logarithmically confined
vortices, or, in the dual language, logarithmically bound dislocations.  Apparently
these dual defects are in the same state at such fixed points.

The duality of the partition function tells us then that there are {\it three} possible
states for the vortices, linearly confined, logarithmically bound, or deconfined --
the last because the symmetry of the effective Hamiltonian ${\cal  H}_{GVM}$
(Eq.~\ref{hgvm}) tells us that if a deconfined dislocation state exists, so must
a deconfined vortex state.   Recent simulation studies~\cite{herb2} have
strongly supported the existence of three such states in the $XY$ model.
For example, if one fixes $K$ below the critical value for the Kosterlitz-Thouless
transition, then at small $h$ one finds fluctuations that scale with system
size $L$ in a way expected for unbound vortices.  At intermediate values of $h$,
the fluctuations behave as if the vortices are in pairs that are interacting logarithmically
at large separations.  For larger values of $h$, they behave like dipoles with linear
confinement~\cite{herb2}.

Finally, one may notice that $\tilde{\cal H}$ is not invariant under $q_x \leftrightarrow q_y$,
so that the fixed points we are examining do not retain the full symmetry
of the underlying lattice.  
Indeed, the loss of symmetry was been apparent in our development of
the RG scaling relations, for which we saw ${\cal H}_0$ and ultimately
${\cal H}_*$ has configurations of vanishing energy along the $q_y=0$
axis, but not on the $q_x=0$ axis.
This problem originated when we went from ${\cal H}_{\rm GVM}$
to our ${\cal H}_{\rm eff}$.  For the former model, the energetics of a dislocation
with a domain wall exiting along a vertical bond is the same for one
in which it exits along a horizontal bond; for the latter, they are
different.  The model can be formulated in a way that retains the lattice
symmetry, at the sacrifice of some of the simplicity of the model we
are currently analyzing.  We will detail how this is done below.
For now we note these models generate unbound dislocation (and
vortex) fixed points that have not one but several nodal lines in
the Brillouin zone, in a way that respects the orientational symmetry
of the lattice -- so that the ${\cal H}_*$ we are 
currently analyzing is really only one of a class of fixed point
Hamiltonians that can arise in $XY$ spin models with symmetry-breaking fields.

\subsection{Integration of Scaling Relations}
\label{scale}
We now need to show that there are parameter regimes for which
${\cal H}_{\rm eff}$ flows to ${\cal H}_*$ both with $\rho=0$ and
$\rho > 0$.  The first two scaling relations -- Eqs.~\ref{scale_rel_1} -- are
trivial to deal with.  The first of these simply states that $K$ is 
invariant as we integrate out short wavelengths: we simply take this
to be a constant in the discussion that follows.  The second is easily
integrated to give
\begin{equation}
\xi^2(\ell)=\xi^2_0e^{-2\ell},
\end{equation}
where $\xi_0=\sqrt{K/h}$ is the bare string width.  Because of the
simple relation between $\xi$ and $\ell$, it is convenient to 
characterize the length scale to which we have integrated in
terms of $\xi$; i.e., $b\equiv e^{\ell} =\xi_0/\xi(\ell)$.  Thus we will
present RG flows below in terms of $\xi$.

Eq.~\ref{scale_rel_2} is also easy to deal with since the parameter $y$
does not renormalize any of the $y_{2n}$'s.   Since the initial value of $\rho$ is
small in our perturbative approach, and it shrinks
under the RG, $y(\ell)$ shrinks rapidly as $\xi(\ell) \rightarrow 0$.
We will not show this explicitly.

Eqs.~\ref{scale_rel_3} by contrast appear formidable since they present
an infinite tower of equations.  However, because the shell integral ${\cal L}$
is the same for each of these, the solutions of the equations are related
to one another by $y_{2n+2} = e^{-2\ell}y_{2n} \equiv \xi^2 y_{2n}/\xi_0^2$, 
as may easily be confirmed by direct substitution.  Recalling $\rho=4\pi^2 y_2$,
this tells us Eq.~\ref{scale_rel_4} may be written as
$$
{{d\rho} \over {d\ell}}=-{2\pi^2{\cal L}\rho \xi^2 \over \xi_0^2}.
$$
Finally, noting $d\xi^2/d\ell=-2\xi_0^2e^{-2\ell}=-2\xi^2(\ell)$, we can rewrite this as
\begin{equation}
{{d\rho} \over {d \xi^2}} = {\pi^2 {\cal L}(\rho,\xi) \rho \over \xi_0^2}.
\label{rg_rho}
\end{equation}
Eq.~\ref{rg_rho} is particularly convenient for integration.  Note the
explicit dependence on $\xi_0^2$ shows that the renormalization
of $\rho$ is strongest for the largest values of $h$: $\rho$ scales
to zero for large values of $h$, leading to the unconfined dislocation
phase.

\begin{figure}[tb]
\begin{center}
 \vbox to 5.0cm {\vss\hbox to 10cm
 {\hss\
   {\includegraphics{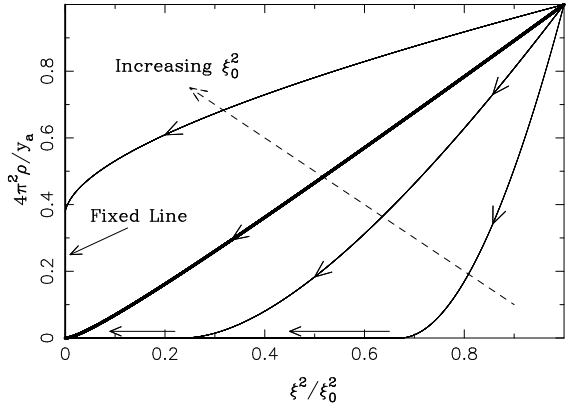}
   }
  \hss}
 }
 \end{center}
\vspace{0.2cm}
\protect \caption{RG flows for Eq.~\ref{rg_rho}.  Left vertical axis is a fixed line, and heavy
line is a separatrix between flows that reach $\rho=0$ for finite $l$ and flow
to the origin as $l \rightarrow \infty$, and those that have $\rho>0$
at the end of their flow.  Finite values of $\rho$ in $H_*$ indicate dislocation binding.}
\protect \label{fig3PRL}
\end{figure}

Fig.~\ref{fig3PRL} shows a typical result for the integration of Eq.~\ref{rg_rho}.
For these flows, we have used our estimate of $y_a$ in terms of 
$h$ to fix the initial value of $\rho$, so that all the flows begin
from the same point in the plot.
The structure that was described in the Introduction is apparent:
there is a fixed line of $\rho$ values, with the $\rho=0$ point -- the
unbound dislocation phase -- an endpoint of this line.  
For small values of $\xi_0$, 
$\rho(\ell)$ lies below the separatrix (shown as a heavy line), 
touches the $\rho=0$ axis
at a finite value of $\xi^2/\xi_0^2 \equiv e^{-2\ell^*}$, and remains on this axis as
it flows to the origin.  
Note that the divergence in ${\cal L}$ as $\rho \rightarrow 0$
guarantees the flows never cross the $\rho=0$ plane to negative
values, which would indicate an instability in our model.
We can interpret $\ell_{scr}=a_0 e^{\ell^*}$ as a
screening length: for separations below this the dislocations appear
to be bound, while for larger separations the interaction is screened by
other dislocations, allowing for an unbinding transition.  Apparently
$\ell_{scr}$
diverges as the transition is approached.
This means the deconfinement transition has a feature in common with
other more standard phase transitions: there is a diverging length scale.
Above the separatrix, the flows end at a non-vanishing value of
$\rho$, which as discussed above indicates the dislocations are in a bound state.
The separatrix represents a deconfinement line, with dislocations
unbound if $\xi_0$ is smaller than some critical value $\xi_{cr}$
for a fixed value of $K$.

The numerical integration of the scaling relations allow us to 
find some critical exponents for how $\rho$ develops and how
$\ell_{scr}$ diverges as the phase boundary is approached.  Before 
giving results for these, we will demonstrate that the lattice
symmetry can be properly incorporated into the model and find the
scaling relations analogous to those above.  This is the subject of the next 
section.

\section{Formulation of the Model with full lattice symmetry }
\label{fullsymmetry}
\subsection{Square Lattice}
\label{SqLatt}
In this section we will formulate the theory retaining the full underlying symmetry of the square lattice.
We begin by defining the bond variables $S_1$ and $S_2$ in terms of integer fields $m_1$ and $m_2$ on the dual lattice sites
$\lbrace {\bf R} \rbrace$ as in Eq.~\ref{mmap} and substitute into Eq.~\ref{ham4} to obtain in momentum space
\begin{eqnarray}
{\cal H}_{\rm VM} &=&{1 \over N}  \sum_{\bf q}\frac 1{2K} 
\left[\bigl|Q_1m_1 ({\bf q})\bigr|^2 + 
\bigl|Q_2m_2 ({\bf q})\bigr|^2 \right]\nonumber \\ 
&+& {1 \over N}\sum_{\bf q} \frac 1{2h} \bigl|Q_1Q_2m_1 ({\bf q}) +
Q_1Q_2m_2 ({\bf q})\bigr|^2,
\label{HVM}
\end{eqnarray}
where $Q_1=1-e^{-i{\bf q \cdot \Delta_1}}$ and $Q_2=1-e^{-i{\bf q \cdot \Delta_2}}$.
As previously, to carry out an RG analysis we need to replace $m_{1,2}$ by continuous fields $\phi_{1,2}$ and add in
cosine terms which tend to preserve the ``integerness'' of the fields, as well their sums and differences. Most such terms
work out to be irrelevant; however, one particular combination turns out to be analog of the $y_a\cos (2\pi a)$ term of
the last section. This is
\begin{eqnarray}
{\cal H}^\prime &=& -y_a\sum_{{\bf r}}\Biggl[\Biggl(1- \cos \biggl\lbrace 2\pi \left[ \phi_1({\bf r+\Delta_2)} - 
\phi_1({\bf r})\right]\biggr \rbrace\Biggr) \nonumber \\
&+& 
\Biggl(1-\cos \biggl\lbrace 2\pi \left[ \phi_2({\bf r+\Delta_1)} - \phi_2({\bf r})\right]
\biggr \rbrace\Biggr)\Biggr].
\label{intsq} 
\end{eqnarray}
It is useful now to do a Fourier transform 
of the integer fields $m_1$ and
$m_2$ in Eq.~\ref{intsq} and expand the result in the small ${\bf q}$ limit
\begin{eqnarray*}
\phi_1({\bf r+\Delta_2)} - \phi_1({\bf r}) &\approx & {1 \over N}\sum_{\bf q} (i{\bf q \cdot \Delta_2})e^{i{\bf q \cdot r}}\phi_1({\bf q}),
\\
\phi_2({\bf r+\Delta_1)} - \phi_2({\bf r}) &\approx & {1 \over N} \sum_{\bf q} (i{\bf q \cdot \Delta_1})e^{i{\bf q \cdot r}}\phi_2({\bf q}). 
\end{eqnarray*}
The effective Hamiltonian is now
\begin{eqnarray}
{\cal H}_{\rm eff} &=& {1 \over N} \sum_{\bf q} \Biggl\lbrace  \frac 1{2K} \left[({\bf q \cdot \Delta_1})^2|\phi_1 ({\bf q})|^2 + 
({\bf q \cdot \Delta_2})^2|\phi_2 ({\bf q})|^2 \right] \nonumber \\ 
&+ &\frac 1{2h} \bigl| ({\bf q \cdot \Delta_1})({\bf q \cdot \Delta_2})\phi_1 ({\bf q}) \cr
&+&({\bf q \cdot \Delta_2})({\bf q \cdot \Delta_1})\phi_2 ({\bf q})\;\;\bigr|^2 \Biggr\rbrace \nonumber \\
&-& y_a\sum_{{\bf r}} \Biggl\lbrace  \left(1- \cos \left[ {{2\pi} \over {N}}
 \sum_{\bf q} (i{\bf q \cdot \Delta_2})e^{i{\bf q \cdot r}}\phi_1({\bf q}) 
 \right] \right) \nonumber \\
&+& \left(1-\cos \left[ {{2\pi} \over {N}} \sum_{\bf q} (i{\bf q \cdot \Delta_1})
e^{i{\bf q \cdot r}}\phi_2({\bf q})\right] \right) \Biggr\rbrace  \nonumber \\
&+& y \sum_{{\bf r}} \Biggl\lbrace  \biggl( 1-\cos [2\pi \phi_1 ({\bf r})] \biggr) \cr
&+& \biggl( 1-\cos [2\pi \phi_2 ({\bf r})] \biggr) \Biggr\rbrace.
\label{Heff}
\end{eqnarray}
In coming to Eq.~\ref{Heff}, we have taken the long wavelength limit $Q_1=i{\bf q \cdot {\bf \Delta}_1}$ and 
$Q_2=i{\bf q \cdot {\bf \Delta}_2}$. We note the symmetry of ${\cal H}_{\rm eff}$ in Eq.~\ref{Heff} under ${\bf {\hat x}}
\leftrightarrow {\bf {\hat y}} $ and $\phi_1 \leftrightarrow \phi_2$ preserves the square symmetry.

We are now set to perform the momentum shell renormalization group procedure on the above effective
Hamiltonian that is perturbative in $y_a$. Note that the $y$ term is irrelevant and shrinks to zero as
discussed in Sec.~\ref{scale}. Our analysis that follows is similar to that described in 
Section \ref{RGanalysis} - we divide the fields $\phi_i$ into long and short wavelength components
$\phi_i^>$ and $\phi_i^<$, integrate out the short wavelength fields (within the momentum shell as shown in
Fig.~\ref{squarefig}) which give an overall prefactor in
the partition function, and finally rescale the fields as in Eq.~\ref{scaling}. In doing this we expand the $y_a$
cosine terms in Eq.~\ref{intsq} and incorporate their Gaussian contribution into our unperturbed Hamiltonian,
which now becomes
\begin{eqnarray}
{\cal H}_{0} = {1 \over N} \sum_{\bf q} &\Biggl[ &\frac 1{2K} \left\{ ({\bf q \cdot \Delta_1})^2 + 
\rho K({\bf q \cdot \Delta_2})^2 \right\}|\phi_1 ({\bf q})|^2  \nonumber \\
&+&  \frac 1{2K} \left\{ ({\bf q \cdot \Delta_2})^2 + 
\rho K ({\bf q \cdot \Delta_1})^2 \right\}|\phi_2 ({\bf q})|^2 \nonumber \\
&+&  \frac {\xi^2}{2K} \Bigl|({\bf q \cdot \Delta_1})({\bf q \cdot \Delta_2})\phi_1 ({\bf q}) \cr
&+& 
({\bf q \cdot \Delta_2})({\bf q \cdot \Delta_1})\phi_2 ({\bf q}) \Bigr|^2 \Biggr],
\end{eqnarray}
where $\rho (\ell) = 4\pi^2y_2(\ell)$ and $\xi(\ell =0) = \sqrt{K/h}$. 
 After rearrangement, the resulting  unperturbed Hamiltonian takes the form,
\begin{eqnarray}
{\cal H}_0 = {1 \over N}  \sum_{\bf q} &\Biggl[ & \frac 1{2K} \Bigl\lbrace ({\bf q \cdot \Delta_1})^2 + 
\rho K ({\bf q \cdot \Delta_2})^2 \cr
&+& \xi^2({\bf q \cdot \Delta_1})^2({\bf q \cdot \Delta_2})^2
\Bigr \rbrace|{\phi_1}({\bf q})|^2  \nonumber \\
&+&  \frac 1{2K}\Bigl \lbrace ({\bf q \cdot \Delta_2})^2 + 
\rho K ({\bf q \cdot \Delta_1})^2 \cr
&+& \xi^2({\bf q \cdot \Delta_2})^2({\bf q \cdot
\Delta_1})^2\Bigr \rbrace |{\phi_2}({\bf q})|^2 \nonumber \\
&+& 
 \frac {\xi^2}{K} ({\bf q \cdot \Delta_1})^2({\bf q \cdot \Delta_2})^2 {\phi_1({\bf q})}{
\phi_2({\bf q})}\Biggr].
\label{ham5}
\end{eqnarray}
As in Sec.~\ref{Scaling_Relation}, the fixed points of the RG equations 
[Eqs.~\ref{scale_rel_1},~\ref{scale_rel_2},~\ref{scale_rel_3}, and~\ref{scale_rel_4}] occur for $y_{2n \geq 4}=0$, and $\xi^2=0$, so that
the generic fixed point Hamiltonian is
\begin{eqnarray}
{\cal H}_* &=&{1 \over N} \sum_{\bf q} \Biggl[  \frac 1{2K} \left\{ ({\bf q \cdot \Delta_1})^2 + 
\rho K ({\bf q \cdot \Delta_2})^2 \right\}|{\phi_1}({\bf q})|^2  \nonumber \\
&+&  \frac 1{2K}\left\{ ({\bf q \cdot \Delta_2})^2 + 
\rho K ({\bf q \cdot \Delta_1})^2 \right\}|{\phi_2}({\bf q})|^2\Biggr].
\label{ham6}
\end{eqnarray}
One may notice that the fixed point Hamiltonian in Eq.~\ref{ham6} is invariant under $q_x
\leftrightarrow q_y$ which implies that the fixed points under investigation do retain the full
underlying symmetry of the lattice. This is different than the case we studied in
Sec.~\ref{bound_state}. As before, the scaling relations show that any small value of
$\rho$ is an attractive fixed point, we identify the $\rho=0$ fixed point as the unbound
dislocation state, and $\rho>0$ as logarithmically bound dislocation states. 
Because of the full lattice symmetry of ${\cal H}_*$ there are now
two nodal lines in the Brillouin zone when $\rho=0$.

For our later calculations we will ignore the cross terms in ${\cal H}_0$  ($\phi_1\phi_2$) as
those terms contribute only at higher orders of $\xi^2$ (starting from $\xi^4$), and the calculation is greatly simplified
because $\langle \phi_1 \phi_2 \rangle_0 =0 $ for the Hamiltonian in Eq.~\ref{ham5} when this is done.
With this simplification,
the propagators for the fields $\phi_1$ and $\phi_2$ calculated from Eq.~\ref{ham5} are
\begin{eqnarray}
&&\langle \phi_1 (-{\bf q}) \phi_1 ({\bf q}) \rangle_0 \cr
&=& \frac 1{({\bf q \cdot \Delta_1})^2 + \rho K ({\bf q \cdot \Delta_2})^2 + 
\xi^2 ({\bf q \cdot \Delta_1})^2({\bf q \cdot \Delta_2})^2 }, \cr
&& \cr
&&\langle \phi_2 (-{\bf q}) \phi_2 ({\bf q}) \rangle_0 \cr
&=& \frac 1{({\bf q \cdot \Delta_2})^2 + \rho K ({\bf q \cdot \Delta_1})^2 + 
\xi^2 ({\bf q \cdot \Delta_1})^2({\bf q \cdot \Delta_2})^2 }.
\end{eqnarray}
With these we define the momentum shell integrals ${\cal L}^i_\Box (\rho, \xi)$ for the the square 
lattice 
\begin{eqnarray}
{\cal L}^1_\Box (\rho, \xi) d\ell &=& K\int^>\!\frac {d^2 {\bf q}}{(2\pi )^2} ({\bf q \cdot \Delta_2})^2 
\langle \phi_1 (-{\bf q}) \phi_1 ({\bf q}) \rangle,
\label{sqint1}\cr
{\cal L}^2_\Box (\rho, \xi) d\ell  &=& K\int^>\!\frac {d^2 {\bf q}}{(2\pi )^2} ({\bf q \cdot \Delta_1})^2
\langle \phi_2 (-{\bf q}) \phi_2 ({\bf q}) \rangle.
\label{sqint2}
\end{eqnarray}
Note ${\cal L}^1_\Box = {\cal L}^2_\Box \equiv {\cal L}_\Box $ because of the square symmetry. The scaling relation 
in Eq.~\ref{rg_rho} remains the same, provided we substitute 
${\cal L} \rightarrow {\cal L}^1_\Box + {\cal L}^2_\Box=2{\cal L}_\Box $. 
The shell integral ${\cal L}_\Box $ may be computed to yield
\begin{eqnarray}
{\cal L}_\Box(\rho, \xi) &=& \frac {4K\Lambda \pi}{\rho K +\xi^2\Lambda^2} \cr
&+& \frac
{4K\Lambda^2}{\sqrt{\rho K (1+\xi^2\Lambda^2)}}\arctan \left(\frac {\pi \sqrt{1+\xi^2\Lambda^2}}
{\Lambda \sqrt{\rho K}}\right)
\nonumber \\
&-&\frac {4K\Lambda^2}{(\rho K+\xi^2\Lambda^2)^{3/2}}\arctan \left(\frac {\pi \sqrt{\rho K+\xi^2\Lambda^2}}{\Lambda}
\right),
\label{LBOX}
\end{eqnarray}
where we have returned to unitless distances, $\Delta_{1,2}=a_0 \equiv 1$.

For small values of $\rho K$ we perform a  Taylor
series to obtain
\begin{equation}
{\cal L}_\Box(\rho, \xi) \approx \frac {K\Lambda^2 }{2 \pi \sqrt{\rho K (1+\xi^2 \Lambda^2)}}.
\label{LBOXshort} 
\end{equation}
We thus see that the scaling relation for $\rho$ in this formulation is identical to Eq. \ref{rg_rho}.
The only caveat is that in our previous formulation, this relation held to all orders
in $\xi^2$, whereas here there are corrections of ${\cal O}(\xi^4)$ and 
higher. In Sec.~\ref{discussion} we will present further results that follow from numerically
integrating the scaling relations.  

\subsection{Triangular Lattice}
\label{TrLatt}
\begin{figure}[tb]
\protect \centerline{\epsfxsize=3in \epsfbox {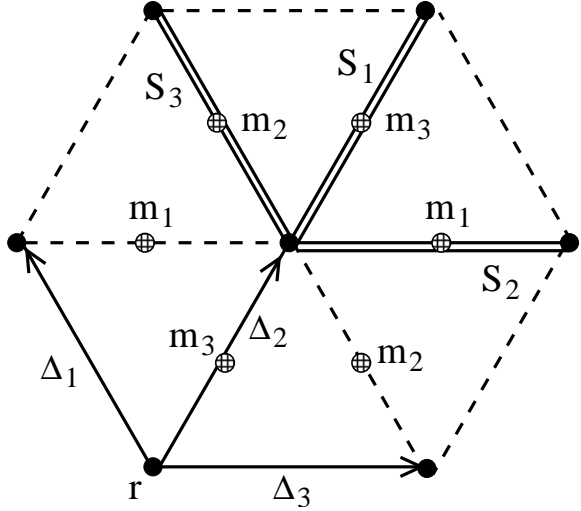}} \vskip .5cm 
\protect \caption{Real lattice sites (${\bf r}$) (black filled circles) and bond center 
 sites (patterned circles) used to represent $S_1,S_2$ and $S_3$ as differences for a triangular lattice. 
 ${\bf \Delta_1}, {\bf \Delta_2}$ and 
${\bf \Delta_3}$ are vectors denoting nearest neighbor bonds. Note that the lines connecting the $m$'s 
cross the bonds they
define, but they are not perpendicular to them.} 
\protect \label{tfig}
\end{figure}

We now consider our theoretical model on a triangular lattice, following the same procedure as in Sec.
\ref{SqLatt}. For our triangular lattice we have three bond variables $S_1,S_2$, and $S_3$ which are defined
via three integer fields $m_1,m_2$ and $m_3$ (respecting the symmetry of the lattice) 
\begin{eqnarray}
S_1 ({\bf r}) &=& m_1({\bf r} + {{\bf \Delta_3}/2}+{\bf \Delta_1}) - m_1 ({\bf r}+ {{\bf \Delta_3}/ 2}),\cr
S_2 ({\bf r}) &=& m_2({\bf r} - {{\bf \Delta_1}/ 2+{\bf \Delta_2}}) - m_2 ({\bf r}- {{\bf \Delta_1}/ 2}), \label{STr} \\
S_3 ({\bf r}) &=& m_3({\bf r} + {{\bf \Delta_2}/2}) - m_3 ({\bf r}+ {{\bf \Delta_2}/2}-{\bf \Delta_3}).\nonumber
\end{eqnarray}
One may note that in Eqs.~\ref{STr} the integer fields $m_{1,2,3}$ are {\it not} defined on the dual lattice sites as 
was done for the square lattice: 
in this case they are  on the centers of the bonds, as illustrated in Fig.~\ref{tfig}.
The lines connecting the $m$'s cross the bonds they define, but they are
not perpendicular to them. The direction of the ${\bf \Delta_i}$'s are also shown in Fig.~\ref{tfig}. Substituting 
Eqs.~\ref{STr} in
Eq.~\ref{ham4} we obtain the unperturbed Hamiltonian in momentum space 
\begin{eqnarray}
&&{\cal H}_{\rm VM}\:\:\:\:\: \cr
&=& {1 \over N} \sum_{\bf q} \Biggl\lbrace \frac 1{2K} 
\left[\bigl|Q_1m_1 ({\bf q})\bigr|^2 + 
\bigl|Q_2m_2 ({\bf q})\bigr|^2 + \bigl|Q_3m_3 ({\bf q})\bigr|^2\right]\nonumber \\ 
&+& \frac 1{2h} \bigl|Q_1Q_2m_1 ({\bf q}) + Q_2Q_3m_2 ({\bf q})+ Q_3Q_1m_3 ({\bf q})\bigr|^2 
\Biggr\rbrace,
\label{HVMTr}
\end{eqnarray}
where $Q_j=1-e^{-i{\bf q \cdot {\bf \Delta_j}}}$. As before we replace the integer fields
$m_{1,2,3}$ with continuous fields $\phi_{1,2,3}$ and add in cosine terms. The 
cosine term that is analogous to $y_a\cos (2\pi a)$ is
\begin{eqnarray}
{\cal H}^\prime &=& -y_a\sum_{{\bf r}}\Biggl[ \Biggl(1- \cos \biggl\lbrace 2\pi \left[ \phi_1({\bf r+\Delta_2)} 
- \phi_1({\bf r})\right]
\biggr \rbrace\Biggr)  \cr
&+&  \Biggl(1-\cos \biggl\lbrace 2\pi \left[ \phi_2({\bf r+\Delta_3)} - \phi_2({\bf r})\right]\biggr \rbrace \Biggr) \cr 
&+& \Biggl(1-\cos \biggl\lbrace 2\pi \left[ \phi_3({\bf r+\Delta_1)} - \phi_3({\bf r})\right] \biggr \rbrace \Biggr)\Biggr]. 
\end{eqnarray} 
We now do a Fourier transform of the fields $m_{1,2,3}$ and expand the result in long wavelength limit
$\phi_i({\bf r+\Delta_j}) - \phi_i({\bf r}) \approx  {1 \over N} \sum_{\bf q} (i{\bf q \cdot \Delta_j})
e^{i{\bf q \cdot r}}\phi_i({\bf q})$ where $i\neq j =1,2,3$.   The resulting effective Hamiltonian is
\begin{eqnarray}
{\cal H}_{\rm eff } &=& {1 \over N} \sum_{\bf q} \Biggl\lbrace  \frac 1{2K} \Biggl[ ({\bf q \cdot
\Delta_1})^2|\phi_1 ({\bf q})|^2 \cr
&+& ({\bf q \cdot
\Delta_2})^2|\phi_2 ({\bf q})|^2 
+ ({\bf q \cdot
\Delta_3})^2|\phi_3 ({\bf q})|^2\Biggr]\nonumber \\ 
&+& \frac 1{2h} \bigl|({\bf q \cdot \Delta_1})({\bf q \cdot \Delta_2})\phi_1 ({\bf q}) \cr 
&+& ({\bf q \cdot \Delta_2})({\bf q \cdot \Delta_3})\phi_2 ({\bf q}) 
+
({\bf q \cdot \Delta_3})({\bf q \cdot \Delta_1})\phi_3 ({\bf q})\;\bigr|^2 \Biggr\rbrace \nonumber \\
- y_a\sum_{{\bf r}}&\Biggl\lbrace & \left( 1- \cos \left[ {{2\pi} \over {N}} \sum_{\bf q} (i{\bf q \cdot \Delta_2})e^{i{\bf q \cdot
r}}\phi_1({\bf q})
\right]\right)  \nonumber \\
&+& \left(  1-\cos \left[{{2\pi} \over {N}} \sum_{\bf q} (i{\bf q \cdot \Delta_3})e^{i{\bf q \cdot
r}}\phi_2({\bf q})\right] \right)\nonumber \\ 
&+& \left( 1-\cos \left[{{2\pi} \over {N}} \sum_{\bf q} (i{\bf q \cdot \Delta_1})e^{i{\bf q \cdot
r}}\phi_3({\bf q}) \right]\right) \Biggr \rbrace \cr
&-& y \sum_{{\bf r}} \Biggl\lbrace  \biggl( 1-\cos [2\pi \phi_1 ({\bf r})] \biggr) \cr 
&+&  \biggl( 1-\cos [2\pi \phi_2 ({\bf r})] \biggr) 
+  \biggl( 1-\cos [2\pi \phi_3 ({\bf r})] \biggr)\Biggr\rbrace .
\label{HeffTr}  
\end{eqnarray}
In deriving Eq.~\ref{HeffTr} we took the long wavelength limit $Q_j = i{\bf q \cdot \Delta_j}$
($j=1,2,3$). We may note that the above effective Hamiltonian is symmetric under interchanges
of the (1,2,3) indices, thus preserving the triangular lattice
symmetry.

Our next step is to do the momentum shell RG on the effective Hamiltonian. As the procedure is 
identical to that discussed in Section~\ref{SqLatt} we present the main results. 
Expanding the cosine terms and absorbing the quadratic term into the unperturbed Hamiltonian,
one finds
\begin{eqnarray}
 {\cal H}_{0} &=& \cr
{1 \over N} \sum_{\bf q} &\Biggl \lbrace &  \frac 1{2K} \Biggl[ \left\{ ({\bf q \cdot \Delta_1})^2 + 
(\rho K)({\bf q \cdot \Delta_2})^2 \right\}|{\phi_1({\bf q})}|^2  \nonumber \\
&+&  \left\{ ({\bf q \cdot \Delta_2})^2 + 
(\rho K)({\bf q \cdot \Delta_3})^2 \right\}|{ \phi_2({\bf q})}|^2 \nonumber \\
&+&  \left\{ ({\bf q \cdot \Delta_3})^2 + 
(\rho K)({\bf q \cdot \Delta_1})^2 \right\}|{\phi_3({\bf q})}|^2 \Biggr]\nonumber \\
+ \frac {\xi^2}{2K}&\bigl|& ({\bf q \cdot \Delta_1})({\bf q \cdot \Delta_2}){\phi_1({\bf q})} \cr
&+& 
({\bf q \cdot \Delta_2})({\bf q \cdot \Delta_3}){\phi_2({\bf q})} \cr
&+& ({\bf q \cdot \Delta_3})({\bf q \cdot \Delta_1}){\phi_3({\bf q})}\;\bigr|^2 \Biggr \rbrace.
\label{H0Tr}
\end{eqnarray}
Rearranging the terms in Eq.~\ref{H0Tr} we get,
\begin{eqnarray}
{\cal H}_0 &=&{1 \over N} \sum_{\bf q} \Biggl \lbrace \frac 1{2K}
\Biggl[ \Bigl\lbrace ({\bf q \cdot \Delta_1})^2 + 
\rho K({\bf q \cdot \Delta_2})^2 \cr
&+& \xi^2({\bf q \cdot \Delta_1})^2({\bf q \cdot \Delta_2})^2
\Bigr\rbrace|{\phi_1}({\bf q})|^2  \nonumber \\
&+&  \Bigl\lbrace ({\bf q \cdot \Delta_2})^2 + 
\rho K ({\bf q \cdot \Delta_3})^2 \cr
&+& \xi^2({\bf q \cdot \Delta_2})^2({\bf q \cdot
\Delta_3})^2\Bigr\rbrace|{\phi_2}({\bf q})|^2 \nonumber \\
&+& \Bigl\lbrace ({\bf q \cdot \Delta_3})^2 + 
\rho K ({\bf q \cdot \Delta_1})^2 \cr
&+& \xi^2({\bf q \cdot \Delta_3})^2({\bf q \cdot
\Delta_1})^2\Bigr\rbrace|{\phi_3}({\bf q})|^2 \Biggr]\nonumber \\
&+& 
 \frac {\xi^2}{K}\Biggl[ ({\bf q \cdot \Delta_1})({\bf q \cdot \Delta_2})^2 ({\bf q \cdot
\Delta_3}){\phi_1({\bf q})}{\phi_2({\bf q})} \nonumber \\
&+& ({\bf q \cdot \Delta_1})({\bf q \cdot \Delta_2})({\bf q \cdot
\Delta_3})^2{\phi_2({\bf q})}{\phi_3({\bf q})} \nonumber \\ 
&+& ({\bf q \cdot \Delta_1})^2({\bf q \cdot \Delta_2})({\bf q \cdot \Delta_3})
{\phi_1({\bf q})}{\phi_3({\bf q})} \Biggr]\Biggr \rbrace.
\label{H02Tr}
\end{eqnarray}
Once again, the generic fixed point Hamiltonian is $y_{2n \geq 4}=0$,  $\xi^2=0$, and
\begin{eqnarray}
{\cal H}_* &=& {1 \over N}\sum_{\bf q} \frac 1{2K} \Biggl[ 
\left\{ ({\bf q \cdot \Delta_1})^2 + 
\rho K({\bf q \cdot \Delta_2})^2 
\right\}|{\phi_1}({\bf q})|^2  \nonumber \\
&+& \left\{ ({\bf q \cdot \Delta_2})^2 + 
\rho K ({\bf q \cdot \Delta_3})^2 
\right\}|{\phi_2}({\bf q})|^2 \nonumber \\
&+&  \left\{ ({\bf q \cdot \Delta_3})^2 + 
\rho K ({\bf q \cdot \Delta_1})^2 
\right\}|{\phi_3}({\bf q})|^2 \Biggr].
\label{fixhamTr}
\end{eqnarray}
Because the fixed point Hamiltonian in Eq.~\ref{fixhamTr} 
captures the triangular symmetry of the lattice,
there are three nodal lines in the Brillouin zone 
for $\rho=0$.  As before we identify
this as the unbound dislocation phase. $\rho>0$ again corresponds to a bound dislocation phase.

\begin{figure}[tb]
\protect \centerline{\epsfxsize=3in \epsfbox {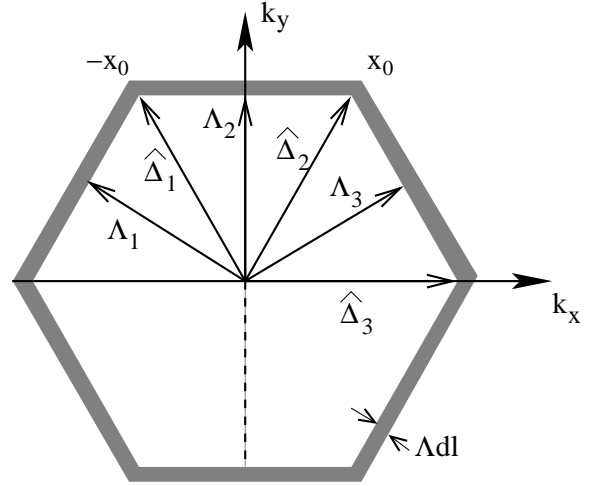}} \vskip .5cm 
\protect \caption{Momentum shell of width $\Lambda d\ell$ in reciprocal space for the triangular lattice.
${\bf \Lambda}_1, {\bf \Lambda}_2$, and ${\bf \Lambda}_3$ are equal length vectors of 
length $2\pi/(\sqrt{3})$. ${\hat \Delta}_1, {\hat \Delta}_2$ and ${\hat \Delta}_3 $ represent 
directions of bond vectors of length $a_0(\equiv 1)$ in Fig.~\ref{tfig}.} 
\protect \label{trianglefig}
\end{figure}

For simplicity of our further calculations we will ignore the cross terms ($\phi_1\phi_2,\phi_2\phi_3,
\phi_1\phi_3$) in ${\cal H}_0$ as those terms contribute only at higher orders of $\xi^2$. In that case 
the propagators of 
Eq.~\ref{H02Tr} can be easily evaluated, yielding
\begin{eqnarray}
\label{triangle_props}
&&\langle \phi_1 (-{\bf q}) \phi_1 ({\bf q}) \rangle_0 \cr
&=& \frac 1{({\bf q \cdot \Delta_1})^2 + \rho K ({\bf q \cdot \Delta_2})^2 + 
\xi^2 ({\bf q \cdot \Delta_1})^2({\bf q \cdot \Delta_2})^2 }, \cr
&&\cr
&&\langle \phi_2 (-{\bf q}) \phi_2 ({\bf q}) \rangle_0 \cr
&=& \frac 1{({\bf q \cdot \Delta_2})^2 + \rho K ({\bf q \cdot \Delta_3})^2 + 
\xi^2 ({\bf q \cdot \Delta_2})^2({\bf q \cdot \Delta_3})^2 }, \\
&&\cr
&&\cr
&&\langle \phi_3 (-{\bf q}) \phi_3 ({\bf q}) \rangle_0 \cr
&=& \frac 1{({\bf q \cdot \Delta_3})^2 + \rho K({\bf q \cdot \Delta_1})^2 + 
\xi^2 ({\bf q \cdot \Delta_3})^2({\bf q \cdot \Delta_1})^2 }.\nonumber 
\end{eqnarray}
The momentum shell integrals that we need to evaluate for the scaling equation are 
\begin{eqnarray}
{\cal L}^1_\triangle d\ell &=& K\int^>\!\frac {d^2 {\bf q}}{(2\pi )^2} 
({\bf q \cdot \Delta_2})^2 \langle \phi_1 (-{\bf q}) \phi_1 ({\bf q}) \rangle,\cr
{\cal L}^2_\triangle d\ell &=& K \int^>\!\frac {d^2 {\bf q}}{(2\pi )^2} 
({\bf q \cdot \Delta_3})^2 \langle \phi_2 (-{\bf q}) \phi_2 ({\bf q}) \rangle,
\label{trint}\\
{\cal L}^3_\triangle d\ell &=& K \int^>\!\frac {d^2 {\bf q}}{(2\pi )^2} 
({\bf q \cdot \Delta_1})^2 \langle \phi_3 (-{\bf q}) \phi_3 ({\bf q}) \rangle.\nonumber
\end{eqnarray}
Because of the lattice symmetry ${\cal L}^1_\triangle= {\cal L}^2_\triangle = {\cal L}^3_\triangle
\equiv {\cal L}_\triangle$. The scaling relation in Eq.~\ref{rg_rho} again remains the same provided we replace 
${\cal L}$ with $3{\cal L}_\triangle$. 
As written,  the integral for ${\cal L}_\triangle$ has to be
done numerically over the momentum shell [shaded portion in Fig.~\ref{trianglefig}].

Alternatively,
an approximate analytical result of the integral can be derived by first doing a Taylor series
expansion with respect to $\xi^2$ in the denominators of Eqs.~\ref{triangle_props}  
and then performing the integration. The result is shown in
Appendix B. For small values of $\rho$, Eqs.~\ref{Ldef},~\ref{L1def},~\ref{L2def} can be expanded 
to obtain an approximate result for 
${\cal L}_\triangle$ which is
\begin{equation}
{\cal L}_\triangle (\rho ,\xi) \approx \frac {\sqrt{3} \Lambda^2 K}{4\pi \sqrt{\rho K}}\left[ 1-\frac 3{8}\xi^2
\Lambda^2 \right].
\label{LTRshort}
\end{equation} 
This result shows that, as before, the shell integral diverges as $\rho \rightarrow 0$.  Once again, the
scaling relation for $\rho$ is identical to Eq.~\ref{rg_rho}, with ${\cal L} \rightarrow 3{\cal L}_\triangle$.

Finally, the evaluation of the constants $\Lambda$ and $x_0$ which define the
shell geometry in Fig.~\ref{trianglefig} are slightly 
involved, so we explicitly show how to evaluate them. We take the real space basis vectors of the 
triangular lattice to be 
${\bf R_1} = a_0 {\bf {\hat x}}$ (along ${\hat \Delta}_3$ in Fig.~\ref{tfig}) and 
${\bf R_2} = -(a_0/2) {\bf {\hat x}} + (\sqrt{3}a_0/2){\bf {\hat y}}$ 
(along ${\hat \Delta}_1$ in Fig.~\ref{tfig}). Then the basis 
vectors of the reciprocal lattice satifying the property ${\bf G_i \cdot R_j} = 2\pi \delta_{ij}$ are
${\bf G_1} = (2\pi/a_0) {\bf {\hat x}} + (2\pi/\sqrt{3}a_0){\bf {\hat y}}$ and 
${\bf G_2} = (4\pi /\sqrt{3}a_0){\bf {\hat y}}$. 
${\Lambda}$ is the distance from the Brillouin zone center to the
middle of one of the edges, and has length equal to half the length
of the reciprocal basis vector ${\bf G_2}$ (See Fig.~\ref{trianglefig}). 
Thus ${\Lambda}=2\pi/(\sqrt{3}a_0)$.  From the momentum shell diagram we 
find that the half-width of the Brillouin zone edge is 
$x_0=\Lambda/\sqrt{3} = 2\pi /(3a_0).$
$\Delta_1, \Delta_2, \Delta_3$ are all equal to the lattice spacing $a_0$ which we have chosen to be of
unit length.

\section{Discussion: Critical Behavior and Phase Diagram}
\label{discussion}

\begin{figure}[tb]
\protect \centerline{\epsfxsize=3in \epsfbox {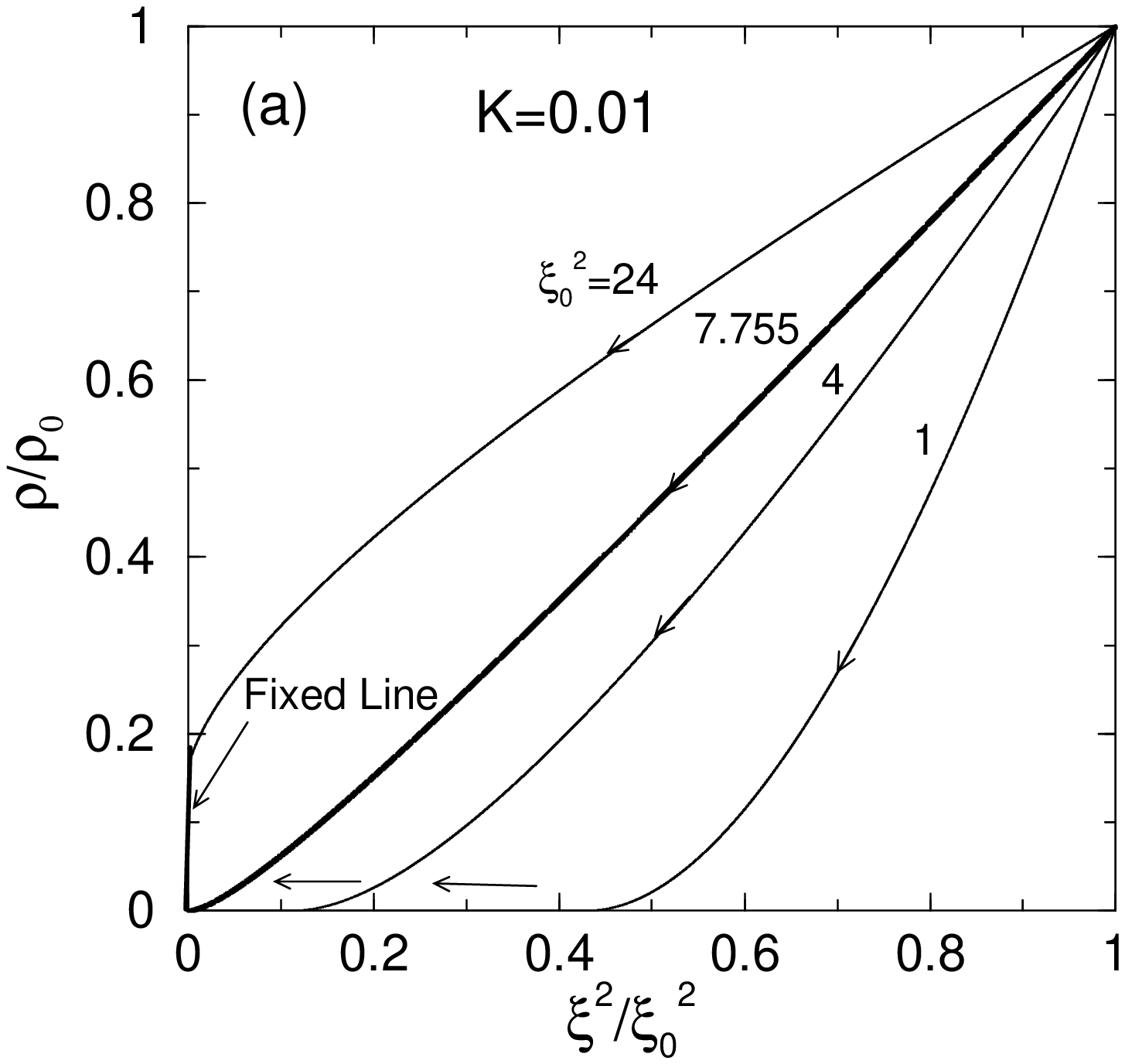}} \vskip .5cm 
\protect \centerline{\epsfxsize=3in \epsfbox {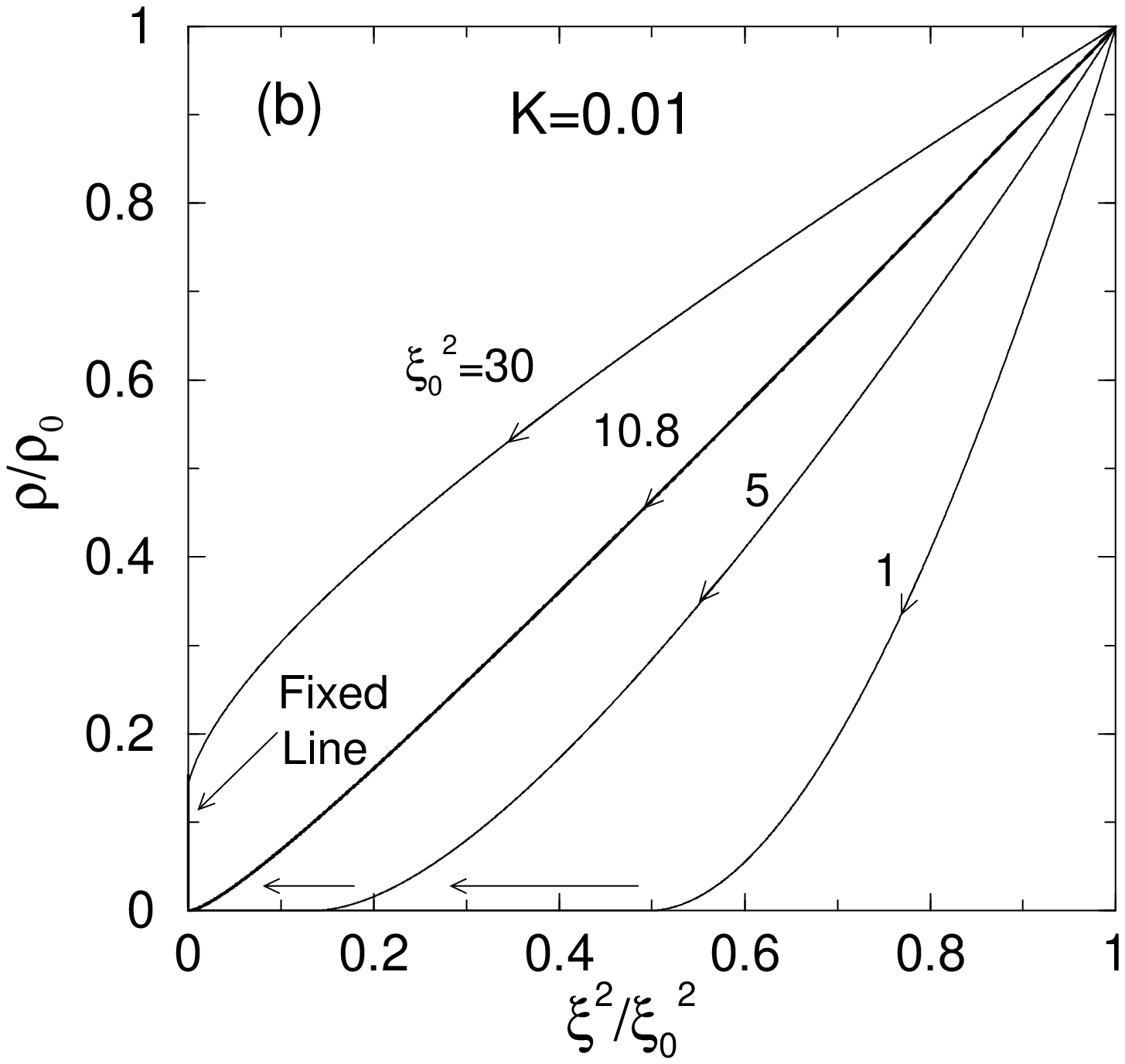}}
\protect \caption{RG flows for the scaling relation in Eq.~\ref{rg_rho} for (a) square lattice 
and (b) triangular lattice with full lattice symmetry. 
$K=0.01$ and initial $\rho_0=1/40$. Left vertical axis is a fixed line, and the thick
line is a separatrix between flows that reach $\rho=0$ for finite $\ell$ and flow to the origin
as $\ell \rightarrow \infty$, and those that have $\rho > 0$ at the end of their flow. 
Thick lines are separatrices with critical values 
of (a) $\xi_{cr}^2=7.755$ and (b) $\xi_{cr}^2=10.8$.} 
\protect \label{RGflow}
\end{figure}

Because the scaling relation for $\rho$ is essentially identical (up to numerical factors)
for the calculations incorporating
the lattice symmetry and for the approximations used in Section~\ref{scale},
it is no surprise that the integration of Eq.~\ref{rg_rho} gives results that are nearly
identical to what we found previously. Fig.~\ref{RGflow} illustrates typical results
for the square and triangular lattice case.  
As before, we find flows that either
accumulate at the $\rho=0$ point, which we identify with the unbound dislocation
phase, or flows that end at $\rho>0$, which represent logarithmically bound
dislocation pairs.  As discussed above, this is a remarkable fixed point structure:
there are no flows whose trajectory change discontinuously as the phase boundary
is crossed; i.e., there are no relevant directions (in the RG sense) leading away 
from the $\rho=0$ fixed point.  One can consider operators other than the ones
we have discussed explicitly -- cosines of higher order derivatives of the $\phi$'s,
analogous to, for example, Eq.~\ref{intsq}-- but it is not hard to convince oneself
that such operators will be more strongly irrelevant than those with which we have worked.
Cosines of integrals of the  $\phi$'s are highly irrelevant as well.

\begin{figure}[tb]
\protect \centerline{\epsfxsize=3.2in \epsfbox {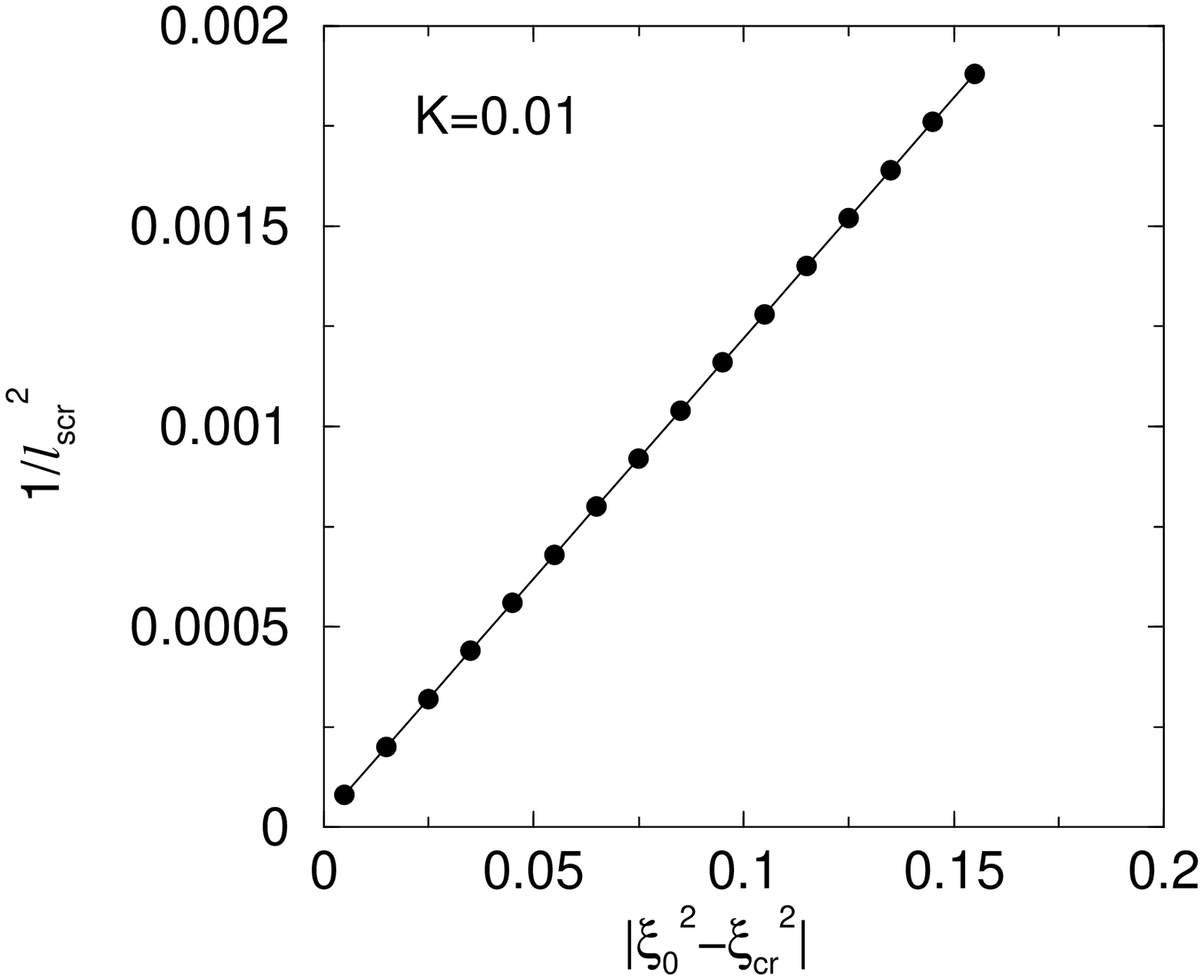}} \vskip .5cm 
\protect \caption{$1/\ell_{scr}^2$ vs. $|\xi^2_0-\xi_{cr}^2|$ for square lattice with full symmetry where
$\ell_{scr}$ is the screening length. $\ell_{scr}$ diverges with a critical 
exponent $-1/2$ as  $\xi^2_0 \rightarrow \xi_{cr}^2 $ from below. Parameters are same as in Fig.~\ref{RGflow}.} 
\protect \label{screen}
\end{figure}

In the absence of any such relevant directions, standard RG theory~\cite{cardy}
tells us that the free energy [$F(\ell \rightarrow \infty)$] is not singular, and
one should not expect to find signals of a phase transition in thermodynamic quantities.  Despite
this, there are some critical properties associated with the deconfinement
transition as we have found it.  
For example, the screening length $\ell_{\rm scr} = a_0 e^{\ell^*}$,
defined by the scale $\ell^*$ at which a flow going to the
unbound dislocation fixed point first strikes the $\rho=0$ axis (discussed in Section~\ref{scale}) ,
diverges as the confinement transition is approached.  Eq.~\ref{rg_rho} suggests
this divergence is a power law in $(\xi_0^2- \xi_{cr}^2)$, which may be
confirmed by examining the flows in detail.  The behavior of $\ell_{\rm scr}$
as a function of $\xi_0^2$ is illustrated in Fig.~\ref{screen}, where it is apparent
that $\ell_{scr}$
diverges with a critical exponent $-1/2$ as  $\xi_0^2 \rightarrow \xi_{cr}^2 $ from below. 
More explicitly, for fixed $K$ this indicates
$\ell_{scr} \sim |h-h_c|^{-1/2}$. The same power law is
obtained for the triangular lattice.

\begin{figure}[tb]
\protect \centerline{\epsfxsize=3.2in \epsfbox {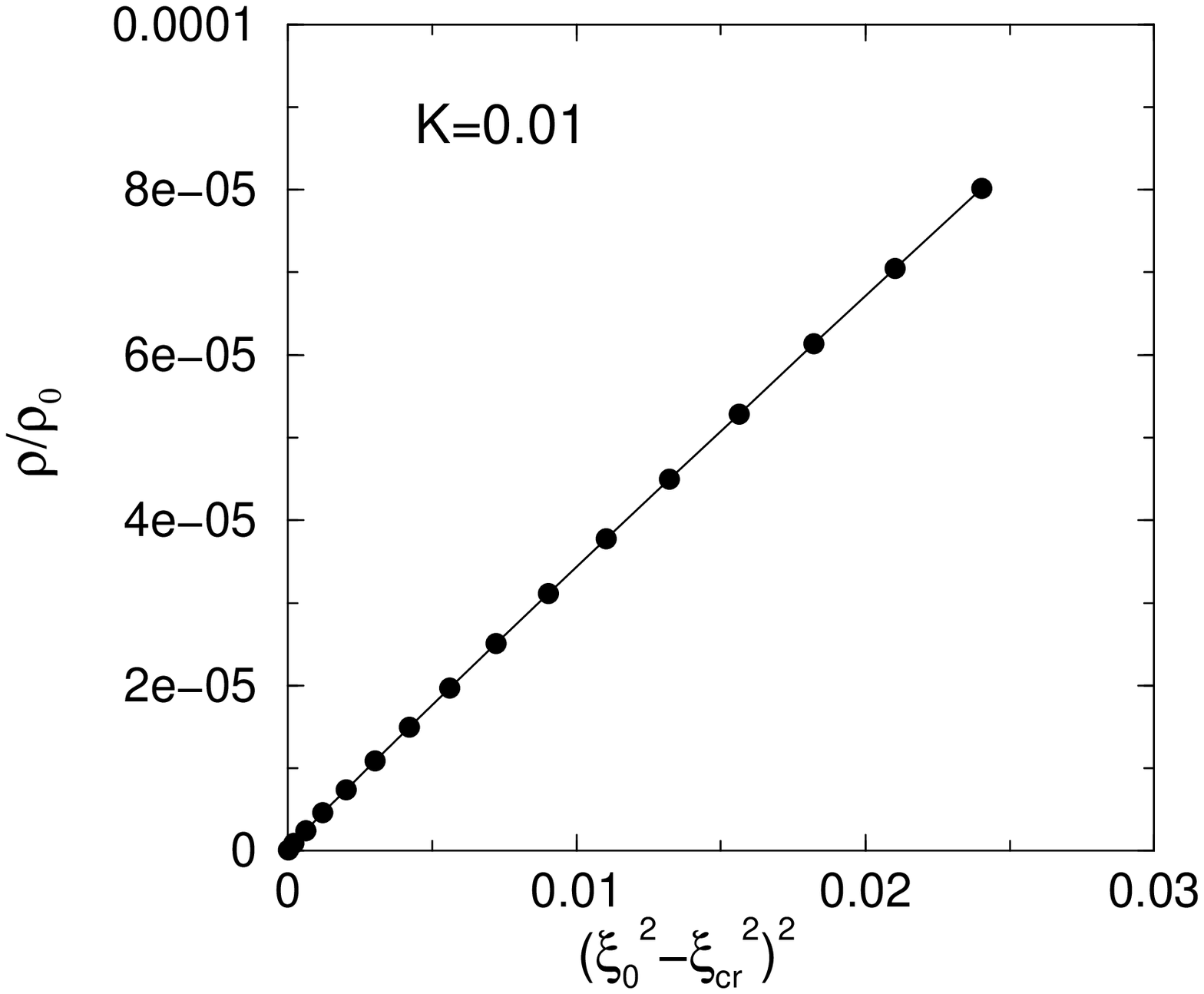}} \vskip .5cm 
\protect \caption{Normalized string tension $\rho/\rho_0$ as a function of 
$(\xi^2_0-\xi_{cr}^2)^2$ for the square lattice with full lattice symmetry.  $\rho$ vanishes quadratically as 
$\xi^2_0 \rightarrow \xi_{cr}^2 $ from above. Parameters are same as in Fig.~\ref{RGflow}. } 
\protect \label{tension}
\end{figure}

Another parameter of interest is the string tension $\rho$, which controls the
effective logarithmic interaction between vortices. Fig.~\ref{tension} illustrates 
how this behaves as $|\xi_0^2-\xi_{cr}^2| \rightarrow 0$.
It is apparent that $\rho$ vanishes quadratically with this difference;
or, equivalently, $\rho \sim |h-h_c|^2$.  In principle the parameter
$\rho$ can be measured if one can create a vortex-antivortex pair at controlled
locations in the system, and measure the force required to separate them.
From the discussion in Section~\ref{bound_state} (see Eq.~\ref{1v}),
it is clear that for a separation $R$, the force required will
have the form $(1+C)/R$, with $C$ diverging when $\rho=0$.
The divergence indicates a change in behavior from a force that falls off
with distance to one that remains finite for any $R$, reflecting the
presence of a string connecting the two vortices.  We note that the diverging
coefficient $C$ does not truly indicate a diverging force; even when $\rho>0$,
for small separations the force will be independent of separation, up to
a characteristic crossover separation $R_c \sim 1/\sqrt{\rho K}$, where the
cutoff in the denominator of Eq.~\ref{1v} becomes apparent and the
$1/R$ force sets in.  Near the transition, the  $(1+C)/R$ behavior only begins
at very large $R$; the effective force between the vortices is thus never large.

Finally, we discuss the reasoning leading to the proposed phase
diagram illustrated in Fig.~\ref{pdsimple}. 
For the purposes of this discussion, we consider $K$ near
its dual point ($K \approx 1/2\pi$), which it should be noted
for $h=0$ is well into the unbound vortex phase.  For large
$h$, the vortices will instead be linearly bound, even at this small value of $K$.
As we know by now,
with increasing $1/h$, there is a transition from an unbound to a 
logarithmically bound dislocation phase.  This may alternatively
be understood as a transition from linearly confined to logarithmically
bound vortices, as described in the last subsection.  Since $K$ is near
the dual point ($K \approx 1/2\pi$), the duality in the model
tells us there must be another transition at large $1/h$ as 
$E_c$ increases from zero, representing a transition from unbound to
logarithmically bound vortices.  The dislocation deconfinement transition
for which we developed the RG analysis
is represented as the large $E_c$ limit of the left
transition line, and its dual is the large $1/h$ limit of the right transition
line.  For vortices, we know that unbinding becomes increasingly
difficult with increasing $h$, so that a smaller  $E_c$ is presumably
required; this is why the phase boundaries move toward the axes
as they approach the origin.  We emphasize that the precise 
behavior of these phase boundaries as $1/h$ and  $E_c$ decrease
is unclear, since our approach requires one (but not both) of these
to be large in order to perform a controlled calculation.  The 
diagram shown in Fig.~\ref{pdsimple} is the simplest that is consistent with
what we have found in our calculations.  As discussed above, the
existence of the three phases is supported by numerical simulation
studies~\cite{herb2}.

\section{Conclusion}
\label{summary}
We have performed a
renormalization group analysis to study vortex unbinding for the 
classical two dimensional $XY$ model in a magnetic field on square and triangular lattices. Our analysis shows
that if one starts at high temperature and large magnetic field, 
vortices in the model unbind as the field is lowered in a two-step process: first strings of overturned spins 
proliferate, and then vortices unbind. 
The proliferated string phase may be understood as one in which the vortices
are logarithmically bound, whereas in the high field phase they are linearly confined.
This transition has an alternate description in terms of domain walls and screw
dislocations, at which the dislocations deconfine.
The vortex deconfinement transition was shown to be dual to this.
Both transitions are remarkably
continuous, but they are not of the Kosterlitz-Thouless type. Finally,  the unbound vortex fixed point 
was shown to
contain a set of nodal lines which are lines of zero energy modes, reflecting the symmetry of the lattice.
\section{Appendix A}
\label{appendixA}
In this Appendix we display the explicit form the shell integral in Eq.~\ref{int_line}.
Defining the parameters $A=1+B$, $B=K\rho+\Lambda \xi^2$, one finds
\begin{eqnarray}
{\cal L}(\rho,\xi)={{2 K \Lambda^2} \over {\pi^2}} &\Biggl\lbrace &\
{{1} \over {\sqrt{A^2-4K\rho \Lambda^2 \xi^2}}} \cr
&\times& \biggl[ {{R_+  - 1} \over {R_+}} \arctan {1 \over {\sqrt{R_+}}} \cr
&-& {{R_-  - 1} \over {R_-}} \arctan {1 \over {\sqrt{R_-}}} \biggr]\cr
&+& {1 \over B} \Biggl[1-{1 \over {\sqrt{AB}}}\arctan \sqrt{B \over A}~ \Biggr]
\Biggr\rbrace.
\label{int_line_exact}
\end{eqnarray}
In Eq.~\ref{int_line_exact}, the parameters $R_{\pm}$ are given by
$$
R_{\pm}=\biggl\lbrace A \pm \sqrt{A^2-4K\rho \Lambda^2 \xi^2} \biggr\rbrace /2\Lambda^2\xi^2.
$$
Eq.~\ref{int_line_exact} can be expanded for small $\rho$ to yield
the result
\begin{eqnarray}
{\cal L}(\rho,\xi)&=&{{2K\Lambda^2} \over {\pi^2}}
\Biggl\lbrace
{{\pi} \over 2} {1 \over {\sqrt{K\rho(1+\xi^2\Lambda^2)}}} + {1 \over {\xi^2\Lambda^2}} \cr
&-&{1 \over {1+\xi^2\Lambda^2}}
+\Biggl[ {1 \over {\Lambda\xi(1+\xi^2\Lambda^2)}}\cr
&-& 
{1 \over {\Lambda^3\xi^3\sqrt{1+\xi^2\Lambda^2}}} \Biggr]
\arctan {{\Lambda\xi} \over {(1+\Lambda^2\xi^2)^{3/2}}}
\Biggr\rbrace \cr
&+& {\cal O}(\sqrt{\rho})
\label{int_line_exp}
\end{eqnarray}
For small values of $\rho$, the first term in Eq.~\ref{int_line_exp} dominates the
RG flows.
\section{Appendix B}
\label{appendixB}
In this Appendix we show the approximate analytical results for the following shell integral 
(See Eqs.~\ref{trint}) with $\mu=\rho K$
\begin{eqnarray}
&& {\cal L}_\triangle (\mu, \xi) d\ell 
= K\int_{\rm shell}\!\frac {d^2 {\bf q}}{(2\pi )^2} \cr
&\times& 
\frac {({\bf q \cdot \Delta_2})^2}{({\bf q \cdot \Delta_1})^2 + \mu ({\bf q \cdot \Delta_2})^2 + 
\xi^2 ({\bf q \cdot \Delta_1})^2({\bf q \cdot \Delta_2})^2 }. 
\end{eqnarray}
To evaluate this integral we expand the denominator for small $\xi^2$. The integral now looks as,
\begin{equation}
{\cal L}_\triangle (\mu, \xi) d\ell = {\cal L}_\triangle^{(a)}(\mu) d\ell+ \xi^2 \frac{d}{d\mu}
{\cal L}_\triangle^{(b)}(\mu) d\ell, 
\label{Ldef}
\end{equation}
where,
\begin{eqnarray}
{\cal L}_\triangle^{(a)}d\ell &=& K \int_{\rm shell}\!\frac {d^2 {\bf q}}{(2\pi )^2} 
\frac {({\bf q \cdot \Delta_2})^2}{({\bf q \cdot \Delta_1})^2 + \mu ({\bf q \cdot \Delta_2})^2}, \\
{\cal L}_\triangle^{(b)} d\ell &=& K \int_{\rm shell}\!\frac {d^2 {\bf q}}{(2\pi )^2} 
\frac {({\bf q \cdot \Delta_1})^2 ({\bf q \cdot \Delta_2})^2}{({\bf q \cdot \Delta_1})^2 + \mu ({\bf q \cdot \Delta_2})^2}.  
\end{eqnarray}
We now display the results of the above integrals.
\begin{eqnarray}
&&{2\pi^2}{\cal L}_\triangle^{(a)} = \frac {8K \Lambda x_0}{1+4\mu} \cr
&-& \frac {4K\sqrt{3} \Lambda^2}{
(1+4\mu)^2}  \ln \left( 
\frac {3\Lambda^2+2\sqrt {3}\Lambda x_0 +x_0^2(1+4\mu)}{3\Lambda^2-2\sqrt {3}\Lambda x_0 +x_0^2(1+4\mu)}\right) \nonumber \\
&+&\frac {2K\sqrt{3}\Lambda^2 (1-4\mu)}{\sqrt{\mu}(1+4\mu)^2}
\arctan \left(\frac{4\sqrt{3}\Lambda x_0\sqrt{\mu}}{3\Lambda^2-x_0^2 (1+4\mu)} \right) \nonumber
\\
&+& \frac {2K \Lambda x_0}{1+\mu} + \frac {2\sqrt{3}K \Lambda^2}{
(1+\mu)^2}  \cr
&\times & \ln \left( 
\frac {3\Lambda^2(1+\mu)+2\sqrt {3}\Lambda x_0(\mu-1) +x_0^2(1+\mu)}{3\Lambda^2(1+\mu)-2\sqrt {3}\Lambda
x_0 (\mu-1)+x_0^2(1+\mu)}\right) \nonumber \\
&+&\frac {2K\sqrt{3}\Lambda^2 (1-\mu)}{\sqrt{\mu}(1+\mu)^2}
\arctan \left(\frac{4\sqrt{3}\Lambda x_0\sqrt{\mu}}{(1+\mu)(3\Lambda^2-x_0^2)} \right) \nonumber
\\
&+& \frac {2K\Lambda x_0}{4+\mu} \cr
&+& \frac {4\sqrt{3} \Lambda^2}{
(4+\mu)^2}  \ln \left( 
\frac {3\Lambda^2 \mu+2\sqrt {3}\Lambda x_0 \mu+x_0^2(4+\mu)}{3\Lambda^2 \mu-2\sqrt {3}\Lambda
x_0 \mu+x_0^2(4+\mu)}\right) \nonumber \\
&-&\frac {2K\sqrt{3}\Lambda^2 (\mu -4)}{\sqrt{\mu}(4+\mu)^2}
\arctan \left(\frac{4\sqrt{3}\Lambda x_0\sqrt{\mu}}{3\Lambda^2 \mu-x_0^2 (4+\mu)} \right).
\label{L1def}
\end{eqnarray}
\begin{eqnarray}
&&{2\pi^2}\frac {d{\cal L}_\triangle^{(b)}}{d\mu} \nonumber \\
&=& -\frac {8K \Lambda \Delta^2 x_0^3}{3(4\mu+1)^2} - 
\frac {24K\Lambda^3 \Delta^2(16\mu^2-32\mu+3) x_0}{(4\mu+1)^4} \nonumber \\
&+&\frac {24\sqrt{3} K\Lambda^4 \Delta^2
(32\mu^2-20\mu+1)}{(4\mu+1)^5} \cr
&\times &\ln \left(\frac {3\Lambda^2+2\sqrt{3}\Lambda x_0+ x_0^2(4\mu+1)}
{3\Lambda^2-2\sqrt{3} \Lambda  x_0+ x_0^2(4\mu+1)} \right)\nonumber \\
&+&\frac {1152K \Lambda^5 \Delta^2 \mu (4\mu-1)x_0^3}{(4\mu+1)^4[[3\Lambda^2+x_0^2(4\mu+1)]^2-12\Lambda^2 x_0^2]} \nonumber \\
&+& \left( \frac {3\sqrt{3}K\Lambda^4 \Delta^2 (192\mu^3-560\mu^2+100\mu-1)}{\sqrt{\mu}(4\mu+1)^5}\right) \cr
&\times& 
\arctan \left(\frac {4\sqrt{3}\Lambda \sqrt{\mu} x_0}{3\Lambda^2-x_0^2(4\mu+1)} \right) \nonumber
\\
&-& \frac {36K\Lambda^5 \Delta^2  [3\Lambda^2+x_0^2(4\mu-1)][16\mu^2-24\mu+1] x_0}{(4\mu+1)^4[9\Lambda^4 +
6\Lambda^2 x_0^2(4\mu-1)+x_0^4(4\mu+1)^2]} \nonumber \\
&-& \frac {K\Lambda \Delta^2 x_0^3}{6(\mu+1)^2} - 
\frac {3K\Lambda^3 \Delta^2 (\mu^2-30\mu+17)x_0}{(\mu+1)^4} \nonumber \\
&-&\frac {12\sqrt{3}K\Lambda^4 \Delta^2 
(2\mu^2-5\mu+1)}{(\mu+1)^5} \cr
&\times & \ln \left(\frac {3\Lambda^2(1+\mu)+2\sqrt{3}\Lambda (\mu-1)x_0 + 
x_0^2(\mu+1)}
{3\Lambda^2 (1+\mu)-2\sqrt{3}\Lambda (\mu-1)x_0 + x_0^2(\mu+1)} \right)\nonumber \\
&+&\frac {288K \Lambda^5 \Delta^2 (3\Lambda^2+x_0^2)x_0\mu (\mu-1)}{(\mu+1)^4[[3\Lambda^2 (1+\mu)+
(\mu+1)x_0^2]^2-12\Lambda^2 x_0^2(\mu-1)^2]} \nonumber \\
&+& \left( \frac {12\sqrt{3}K\Lambda^4 \Delta^2 (3\mu^3-35\mu^2+25\mu-1)}{\sqrt{\mu}(\mu+1)^5}\right) \cr
&\times & \arctan \left(\frac {4\sqrt{3}\Lambda \sqrt{\mu} x_0}{(1+\mu)(3\Lambda^2-x_0^2)} \right) \nonumber
\\
&+& \Bigl\lbrace 144K\Lambda^5 \Delta^2 (3\Lambda^2-x_0^2)x_0 (\mu-1)(\mu^2-6\mu+1)\Bigr\rbrace \cr
&\times& \Bigl\lbrace (\mu+1)^4\Bigl[9 \Lambda^4 -
6\Lambda^2 x_0^2+x_0^4(\mu+1)^2 +18\Lambda^4\mu \cr
&+& 36 \Lambda^2x_0^2 \mu
-6\Lambda^2x_0^2\mu^2+9\Lambda^4 \mu^2\Bigr]\Bigr\rbrace^{-1} \nonumber \\
&-&\frac {2K\Lambda \Delta^2 x_0^3}{3(\mu+4)^2} + 
\frac {144K\Lambda^3 \Delta^2 (\mu-4) x_0}{(\mu+4)^4} \nonumber \\
&-&\frac {48\sqrt{3}K\Lambda^4 \Delta^2 
(\mu^2-10\mu+8)}{(\mu+4)^5} \cr
&\times & \ln \left(\frac {3\Lambda^2 \mu+2\sqrt{3}\Lambda \mu x_0+ x_0^2(\mu+4)}
{3\Lambda^2 \mu-2\sqrt{3}\Lambda \mu x_0+ x_0^2(\mu+4)} \right)\nonumber \\
&+&\frac {1152K\Lambda^5 \Delta^2\mu (\mu-4)x_0^3}{(\mu+4)^4[[3\Lambda^2
\mu+(\mu+4)x_0^2]^2-12\Lambda^2 \mu^2 x_0^2]} \nonumber \\
&+& \left( \frac {3\sqrt{3}K\Lambda^4 \Delta^2(3\mu^3-140\mu^2+400\mu-64)}{\sqrt{\mu}(\mu+4)^5}\right) \cr
&\times & \arctan \left(\frac {4\sqrt{3}\Lambda \sqrt{\mu} x_0}{3\Lambda^2 \mu-x_0^2(\mu+4)} \right) \nonumber
\\
&+& \Bigl \lbrace 36K\Lambda^5 \Delta^2 [3\Lambda^2 \mu+x_0^2(4-\mu)]x_0[\mu^2-24\mu+16]\Bigr \rbrace \cr
&\times & \Bigl \lbrace (\mu+4)^4\Bigl[9\Lambda^4
\mu^2 +24\Lambda^2 \mu x_0^2- 6\Lambda^2\mu^2 x_0^2 \cr
&+& x_0^4(\mu+4)^2\Bigr]\Bigr \rbrace^{-1}.
\label{L2def}
\end{eqnarray}

\acknowledgements

The authors would like to thank many colleagues for helpful discussions and
suggestions, particularly Joseph P. Straley, Ganpathy Murthy, Donald Priour,
Andy Lau, and Ramin Golestanian.  This work was supported by NSF Grant
No. DMR-0108451.


\begin{thebibliography}{11}
\bibitem{com2} In practice we will not need to enforce the boundary
constraint, since relaxing it does not lead to spurious divergences in
our RG analysis.
\end{thebibliography}
\end{document}